\documentclass[superscriptaddress,nobibnotes,amsmath,amssymb,notitlepage,twocolumn,prb,longbibliography]{revtex4-2}

\usepackage{bm,braket}
\usepackage[toc,page]{appendix}
\usepackage{comment}
\usepackage{dcolumn}
\usepackage{wasysym}
\usepackage{amsfonts}

\usepackage{graphicx,color,hyperref}
\usepackage[caption=false]{subfig}
\hypersetup{colorlinks=true, linkcolor=blue, citecolor=blue, urlcolor=blue} 

\graphicspath{{./img/}}

\newcommand{\beq}[1]{\begin{equation}\label{#1}}
\newcommand{\eep}{\;.\end{equation}}
\newcommand{\eec}{\;,\end{equation}}
\newcommand{\eeq}{\end{equation}}

\newcommand*\dd{\mathop{}\!\mathrm{d}} 

\newcommand{\om}{\omega}

\newcommand{\Om}{\Omega}

\DeclareMathAlphabet{\mathcal}{OMS}{cmsy}{m}{n} 

\renewcommand{\vec}[1]{{\bf #1}}

\newcommand{\kv}{\vec{k}}

\newcommand{\rv}{\vec{r}}

\usepackage{amsmath}
\usepackage{amssymb}
\usepackage{xcolor}
\usepackage{bbm}
\usepackage{physics}
\usepackage{float}
\usepackage{graphicx}
\usepackage{dcolumn} 
\usepackage{bm} 
\usepackage{siunitx}
\usepackage{enumitem}  

\makeatletter
\renewcommand*{\fnum@figure}{{\normalfont\bfseries \figurename~\thefigure}}
\makeatother

\allowdisplaybreaks

\definecolor{orange}{rgb}{1,0.5,0}

\graphicspath{{./img/}}

\DeclareMathAlphabet{\mathcal}{OMS}{cmsy}{m}{n} 

\newcommand{\intBZ}{\int_{\text{BZ}}} 

\makeatletter
\newcommand{\specificthanks}[1]{\@fnsymbol{#1}}
\makeatother

\begin{document}

\preprint{APS/123-QED}

\title{Anomalous geometric transport signatures of topological Euler class}

\newcommand{\TCM}{{Theory of Condensed Matter Group, Cavendish Laboratory, University of Cambridge, J.\,J.\,Thomson Avenue, Cambridge CB3 0HE, UK}}
\newcommand{\OXmath}{Mathematical Institute, University of Oxford,
Andrew Wiles Building, Woodstock Road, Oxford, OX2 6GG, UK}
\newcommand{\UoM}{Department of Physics and Astronomy, University of Manchester, Oxford Road, Manchester M13 9PL, UK}

\author{Ashwat Jain}
\email{jain.ashwat@gmail.com}
\thanks{Contributed equally.}
\affiliation{\OXmath}

\author{Wojciech J. Jankowski}
\email{wjj25@cam.ac.uk}
\thanks{Contributed equally.}
\affiliation{\TCM}

\author{Robert-Jan Slager}
\email{rjs269@cam.ac.uk}
\thanks{}
\affiliation{\UoM}
\affiliation{\TCM}

\date{\today}
\begin{abstract}
    We investigate Riemannian quantum-geometric structures in semiclassical transport features of two-dimensional multigap topological phases. In particular, we study nonlinear Hall-like bulk electric current responses and, accordingly, semiclassical equations of motion induced by the presence of a topological Euler invariant. We provide analytic understanding of these quantities by phrasing them in terms of momentum-space geodesics and geodesic deviation equations and further corroborate these insights with numerical solutions. Within this framework, we moreover uncover anomalous bulk dynamics associated with the second- and third-order nonlinear Hall conductivities induced by a patch Euler invariant. As a main finding, our results show how one can reconstruct the Euler invariant by coupling to electric fields at nonlinear order and from the gradients of the electric fields. Furthermore, we comment on the possibility of deducing the nontrivial non-Abelian Euler class invariant specifically in second-order nonlinear ballistic conductance measurements within a triple-contact setup, which was recently proposed to probe the Euler characteristics of more general Fermi surfaces. Generally, our results provide a route for deducing the topology in real materials that exhibit the Euler invariant by analyzing bulk electrical currents.    
    
\end{abstract}

\maketitle

\section{Introduction} Anomalous electrodynamic responses constitute a key experimental signature of topological phases of matter~\cite{Rmp1, Rmp2, Weylrmp}. These responses are related to field-theoretic anomalies realized in topological vacua~\cite{Arouca2022}. In two spatial dimensions, quantum Hall effects, for example, are manifested by transverse dissipationless currents. Consequently, the appearance of this effect in the context of Chern insulators can be understood as a consequence of the parity anomaly~\cite{haldane1988, Fradkin1986}, as further supported by the Chern-Simons theories. In three dimensions, topological $\mathbb{Z}_2$ insulators and axion insulators, recently shown to exist in manganese-doped bismuth telluride compounds~\cite{otrokov2019prediction, jo2020}, can realize topological $\theta$~terms~\cite{Zhang2008, Zhang2009}, which in turn lead to an anomalous topological magnetoelectric effect. Similarly, Weyl semimetals are associated with chiral anomalies~\cite{Kharzeev_2014, Sid2014, Behrends_2016},
and are understood in terms of the dynamics of axial gauge fields emergent from nontrivial band topology~\cite{Vazifeh2013}. Moreover, apart from these electromagnetic anomalies, topological matter has been shown to exhibit an interplay with background geometry, as realized through gravitational-axial anomalies in Weyl semimetals~\cite{Cortijo2015, Gooth2017}, or nontrivial torsional responses~\cite{Codefects2, Mode1, Parrikar_2014, Ferreiros2019, jaakko2020, Nissinen_2023}. Gravitational anomalies in topological materials~\cite{Stone2012} open new avenues for exotic phenomena associated with Hall viscosity and elastic responses~\cite{Volovik2019}. 

The aforementioned anomalies are by and large associated with single gap and gapless topologies, which are extensively classified~\cite{Po_2017, Slager_2013,Kruthoff_2017,Bradlyn_2017,Shiozaki14, SchnyderClass}. In contrast, the anomalous electrodynamic responses of more recently introduced systems with multiple topological band gaps~\cite{bouhon2020geometric, davoyan2024, jankowskiPRB2024disorder, jankowskiPRB2024Hopf}, which require finite band partitionings to culminate in multiband invariants classified via homotopy theory, remain rather unexplored, although the first quantized optical responses therein were recently identified~\cite{jankowski2023optical, jankowskiPRL2024}. A~paradigmatic multigap topological invariant is the Euler class~\cite{bouhon2018wilson, BJY_nielsen, bouhon2019nonabelian, bouhon2020geometric} protected by $\mathcal{C}_2 \mathcal{T}$ (i.e. two-fold rotation combined with time-reversal) symmetry, which can be associated with the non-Abelian nodal charges and their braiding processes between different gaps. That is, because of the anticommutative nature of band node charges in such systems~\cite{doi:10.1126/science.aau8740}, braiding them in momentum space can result in a set of two bands with similarly, rather than oppositely, charged nodes. Similar charges prevent the nodes from annihilating, which is encapsulated by a nonzero Euler class that can only change upon braiding with a node in an adjacent band gap~\cite{BJY_nielsen, bouhon2019nonabelian}. 

\begin{figure}[t!]
\centering
\includegraphics[width=\columnwidth]{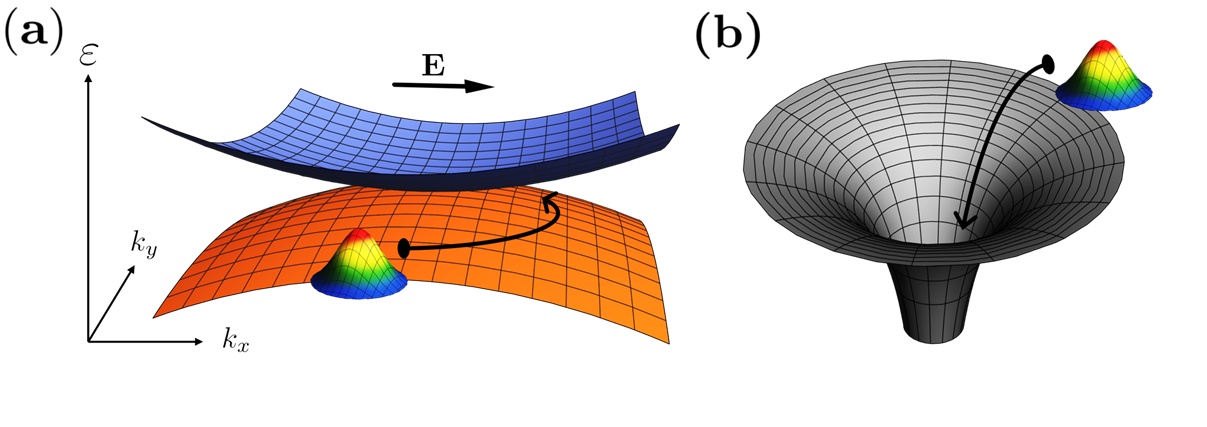}
  \caption{(\textbf{a}) Non-Abelian band singularity hosting a nontrivial patch Euler class $\chi$ and realizing anomalous wavepacket dynamics. \textbf{(b)} Associated momentum-space quantum metric singularity owing to topologically nontrivial Euler node. The Riemannian geometry is fingerprinted through the momentum-space Christoffel symbols and Riemann curvature tensor associated with the singularity in the quantum metric induces anomalous wavepacket dynamics and nonlinear bulk current responses allowing a reconstruction of the Euler invariant.}
\label{Fig:Intro}
\end{figure}

The absence of an anomalous transport signature of Euler topology is an open problem, which we address in this work.
The Euler invariant $\chi \in \mathbb{Z}$ defined for a pair of adjacent bands $\ket{u_{n}}, \ket{u_{n+1}}$ reads~\cite{bouhon2018wilson, BJY_nielsen, bouhon2019nonabelian, bouhon2020geometric}
\beq{eq:eq1}
    \chi = \frac{1}{2\pi} \int_\mathcal{D} \text{Eu} - \frac{1}{2\pi} \int_\mathcal{\partial D} a
\eec
with the Euler curvature $\text{Eu} = \text{d}a$ defined from the non-Abelian Euler connection $a = \bra{u_n} \ket{\text{d} u_{n+1}}$ over a Brillouin zone (BZ) patch $\mathcal{D} \in \text{BZ}$~\cite{bouhon2019nonabelian}. Non-Abelian Euler topologies appear elusive to dc transport signatures, despite the fact that they are established in quantum simulators~\cite{unal2020, zhao2022observation, breach2024interferometry}, through metamaterial realizations~\cite{Jiang2021, jiang_meron}, phonon bands~\cite{Peng2021, Peng2022Multi}, in out-of-equilibrium phases~\cite{slager2022floquet,unal2020}, are predicted in real material/magnetic systems~\cite{bouhon2019nonabelian, Konye2021, lee2024}, and more recently, in interacting contexts~\cite{wahl2024}. In particular, under $\mathcal{C}_2 \mathcal{T}$ symmetry, the Berry curvature in the occupied bands vanishes identically. Hence, the standard anomalous Hall velocity of electrons vanishes in Euler phases, whereas the entire topology-induced geometric transport can be captured by a quantum metric, which is real and positive-semidefinite~\cite{provost1980riemannian, Ahn2020, Ahn2021, bouhon2023quantumgeometry}. In particular, any band node associated with the Euler invariant can be viewed as a quantum-metric singularity. Such a singularity can in some context be thought of as a momentum-space black hole, in the sense that its momentum-space Kretschmann invariant~\cite{K1915}, i.e., a scalar constructed from the contraction of the Riemann curvature tensor with itself, diverges at the band node; see Fig.~\ref{Fig:Intro}.

The Euler invariant can be viewed as a winding of an interband dipole moment, in direct contrast to the winding of the ground-state electric polarization that defines the well-established Chern invariant. As a consequence, optical effects in response to ac electric fields can be quantized. Correspondingly, it was shown that the Euler invariant can induce a quantized optical conductivity and the ratios of third-order jerk photoconductivities~\cite{Fregoso2018} quantized by the invariant~\cite{jankowski2023optical}, by leveraging the associated singular quantum-geometric conditions~\cite{bouhon2023quantumgeometry}. Given the interplay of ac and dc responses, as underpinned by the Kramers-Kronig relations, this provides further possibilities that the multiband invariant can be extracted from the responses to static electric fields~\cite{jankowski2023optical}, especially as the multiband contributions are known to emerge in the field-induced corrections of the nonlinear transport~\cite{Gao2014}.

In this work, we study anomalous transport signatures associated with the quantum geometry of the non-Abelian bands hosting a multigap Euler topology. We show that, despite the vanishing Hall responses at linear order, there are enhanced nonlinear geometric responses in dc transport present, which we retrieve from the equations of motion of individual wavepackets. While in centrosymmetric Euler phases, the combined second-order bulk responses need to vanish by the inversion symmetry, the noncentrosymmetric Euler bands realize the second-order quantum-metric nonlinear Hall effect~\cite{Gao2023} and responses to electric-field gradients induced by the quantum-metric dipole~\cite{Lapa2019}. Finally, and most importantly, we show that the Euler bands universally exhibit a third-order Hall response~\cite{Lai2021, Mi2023, Gao_2023_2, Mandal2024, Li2024} in the nondegenerate limit, and hence that Euler bands can realize higher-order dissipationless currents. As a consequence, we showcase bulk observable anomalous dc transport signatures of nodal Euler band topology.

\section{Multiband Riemannian quantum geometry}\label{sec:II} To retrieve the multiband quantum-geometric electric transport responses induced by the Euler invariant, we define the momentum-space geometric background induced by the topological bands. We begin with the definition of the multiband quantum metric~\cite{provost1980riemannian, Ahn2020, Ahn2021, bouhon2023quantumgeometry, jankowski2023optical} realized by bands $n,m$
\beq{}
    g^{nm}_{\alpha \beta} = \frac{1}{2} \Big[ \bra{\partial_{k_\alpha} u_{n\kv}} \ket{u_{m\kv}} \bra{u_{m\kv}} \ket{\partial_{k_\beta} u_{n\kv}} + \text{c.c.} \Big]
\eec
which is manifestly real and symmetric, and defines a quantum distance $ds^2_{nm} = g^{nm}_{\alpha \beta} \text{d}k_\alpha \text{d}k_\beta$, where we assumed Einstein summation convention, as in the rest of this work. Furthermore, we define a combined band-weighted quantum metric as
\beq{}
    \mathcal{G}^n_{\alpha \beta} = \sum_{\{m\}} c_{nm} g^{nm}_{\alpha \beta},
\eeq
where one can restrict the sum to involve all the other bands $m \neq n$, or otherwise, specifically the unoccupied bands $(\text{unocc})$~\cite{Ahn2020, Ahn2021, bouhon2023quantumgeometry, jankowski2023optical}. Notably, setting ${c_{nm} = 1}$ for~unoccupied bands ($m$) retrieves the total quantum metric in the occupied bands $g^n_{\alpha \beta} = \sum^{\text{unocc}}_m g^{nm}_{\alpha \beta}$. Physically, $g^n_{\alpha \beta}$ is related to the electric quadrupole moment in the bands and hence couples to the nonuniform electric fields through its derivatives~\cite{Lapa2019, Souza2023}. On the contrary, setting the weighting coefficients to band energy weights ${c_{nm} = \hbar/(\varepsilon_{n\kv}-\varepsilon_{m\kv})}$, with $\varepsilon_{n\kv}, \varepsilon_{m\kv}$ the energies of bands $n,m$, retrieves a weighted quantum-metric ${G^n_{\alpha \beta} = \hbar~\sum_{m \neq n} g^{nm}_{\alpha \beta}/(\varepsilon_{n\kv}-\varepsilon_{m\kv})}$, which through its momentum-space multipoles enters the nonlinear responses to static uniform electric fields~\cite{Gao2014}. The generalized quantum metric likewise defines a renormalized momentum-space background geometry ${ds^2_{n}(\{ c_{nm} \}) = \mathcal{G}^n_{\alpha \beta} \text{d}k_\alpha \text{d}k_\beta}$. In this work, we utilize such weighted quantum geometries probed by both types of couplings to electric fields, i.e., $\mathcal{G}^n_{\alpha \beta}(\nabla)$ for nonuniform electric fields, and $\mathcal{G}^n_{\alpha \beta}(E^2)$ for quadratic response in electric field strengths. In particular, a band node with nontrivial Euler invariant provides for a singularity in both considered types of weighted quantum metrics.

Furthermore, we define the Christoffel symbols for the connections in the weighted quantum metrics,
\beq{}
    \Gamma^n_{\alpha \beta \gamma} = \frac{1}{2} (\partial_{k_\gamma} \mathcal{G}^n_{\alpha \beta} + \partial_{k_\beta} \mathcal{G}^n_{\alpha \gamma} - \partial_{k_\alpha } \mathcal{G}^n_{\beta \gamma})
\eec
and the weighted Riemannian curvature tensor ${\mathcal{R}^n_{\alpha \beta \gamma \delta} = \partial_{k_\gamma} \Gamma^n_{\alpha \beta \delta} - \partial_{k_\delta} \Gamma^n_{\alpha \beta \gamma} + [\vec{\Gamma}^n_{\alpha \delta}, \vec{\Gamma}^n_{\beta \gamma}]}$. Additionally, under the vanishing commutation of the second term, $[\vec{\Gamma}^n_{\alpha \delta}, \vec{\Gamma}^n_{\beta \gamma}] \equiv \Gamma^n_{\mu \alpha \delta}  \Gamma^n_{\beta \mu \gamma} - \Gamma^n_{\nu \beta \gamma} \Gamma^n_{\alpha \nu \delta} = 0$  (see Appendix~\ref{app:B}), the weighted Riemannian curvature tensor reduces in terms of the quantum metric to
\begin{small}
\beq{}
  \mathcal{R}^n_{\alpha \beta \gamma \delta} = \frac{1}{2}(\partial_{k_\alpha} \partial_{k_\delta} \mathcal{G}^n_{\beta \gamma} + \partial_{k_\beta} \partial_{k_\gamma} \mathcal{G}^n_{\alpha \delta} - \partial_{k_\alpha} \partial_{k_\gamma} \mathcal{G}^n_{\beta \delta} - \partial_{k_\beta} \partial_{k_\delta} \mathcal{G}^n_{\alpha \gamma}).
\eeq
\end{small}
These Riemannian-geometric quantities are central objects for the wavepacket dynamics and the Euler-invariant reconstruction scheme, as well as for the bulk second- and third-order anomalous dc responses in the next and the following sections, respectively.
\begin{figure*}[t!]
\centering
\includegraphics[width=0.8\linewidth]{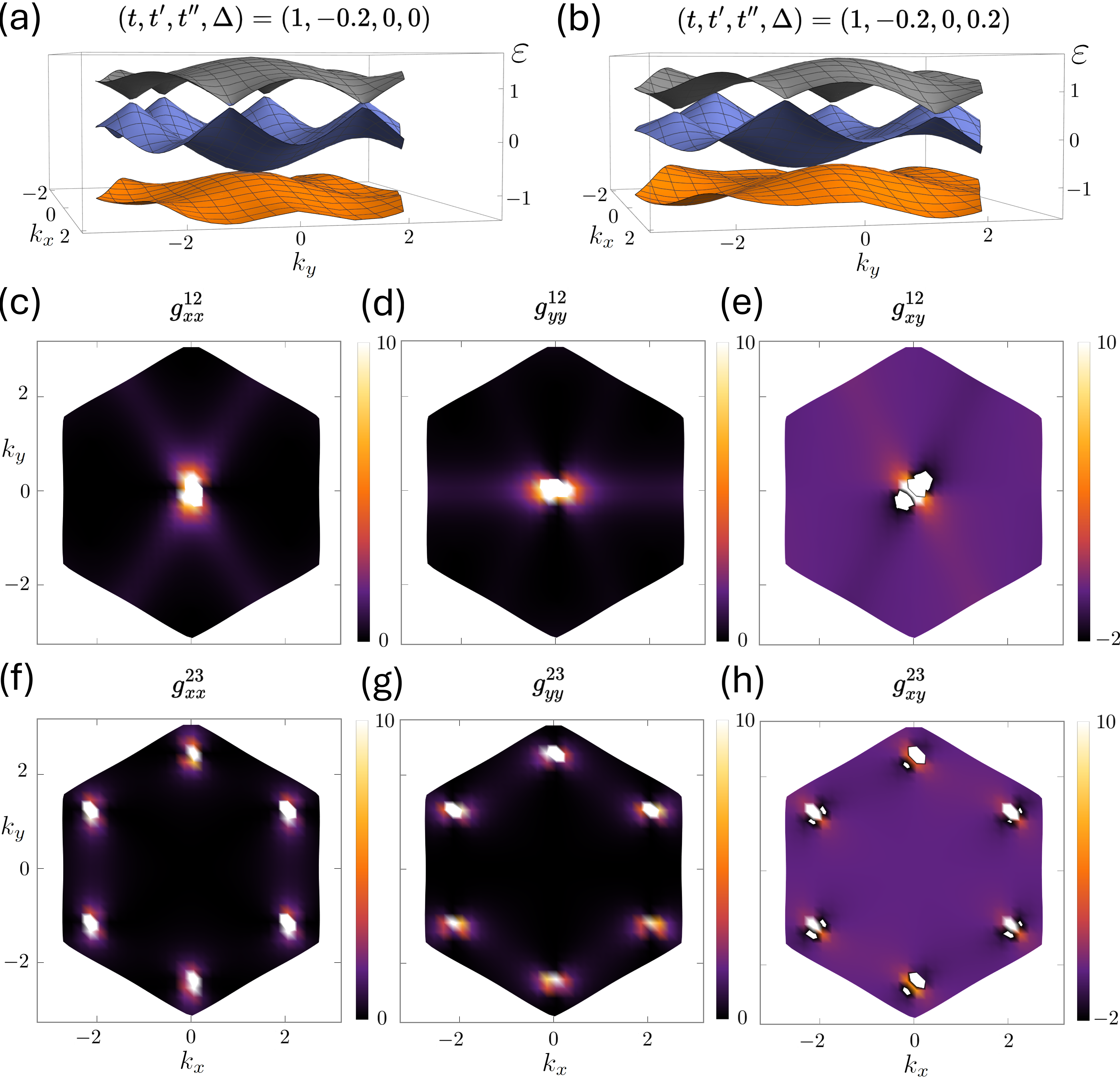}
  \caption{Lattice-regularized model with band nodes exhibiting patch Euler class $\chi = 1$. Band structure in the \textbf{(a)} centrosymmetric ($\Delta=0$) and \textbf{(b)} noncentrosymmetric ($\Delta = 0.2$) model realizations. \textbf{(c)}--\textbf{(h)} Momentum-space geometries captured by the multiband quantum metrics $g^{nm}_{ij}$ realized between the bands supporting nontrivial patch Euler class $n=1,2$, and the neighboring bands $m$ for different spatial components $i,j=x,y$: \textbf{(c)} $g^{12}_{xx}$, \textbf{(d)} $g^{12}_{yy}$, \textbf{(e)} $g^{12}_{xy}=g^{12}_{yx}$, and \textbf{(f)} $g^{23}_{xx}$, \textbf{(g)} $g^{23}_{yy}$, \textbf{(h)}~$g^{23}_{xy}=g^{23}_{yx}$. We note that the inversion symmetry-breaking parameter $\Delta=0.2$ introduces asymmetry in the band energies. The metric components $g^{12}_{ij}$ and $g^{23}_{ij}$ become singular at the Brillouin zone center and corners, respectively, which is consistent with the momentum-space position of the band degeneracies.}
\label{Fig:Model}
\end{figure*}

\section{Model}\label{sec:III} While the introduced geometric framework is general and extends beyond particular models, to pinpoint the momentum-space Riemannian geometry associated with the Euler class, we employ a lattice-regularized model with patch Euler class $\chi = 1$~\cite{Jiang2021, wahl2024} on a kagome lattice. The model Hamiltonian reads
\\
\begin{small}
\beq{}
    H = \sum_{i} \varepsilon_i c^\dagger_{i} c_{i} - t \sum_{\langle i, j \rangle}  c^\dagger_{i} c_{j} - t' \sum_{\langle \langle i, j \rangle \rangle}  
 c^\dagger_{i} c_{j} -t'' \sum_{\substack{\langle \langle \langle i, j \rangle \rangle \rangle_{\hexagon}}}  c^\dagger_{i} c_{j},
\eeq
\end{small}
where $i,j$ run over orbitals $A, B, C$, where single-particle annihilation/creation operators are denoted as $c_i$/$c^\dagger_i$, where $\varepsilon_i$ are onsite energies, and where $t$, $t'$, $t''$ are nearest neighbor ($\langle i, j \rangle$), next-nearest neighbor ($\langle \langle i, j \rangle \rangle$), and next-next nearest neighbor hoppings across plaquettes (${\langle \langle \langle i, j \rangle \rangle \rangle_{\hexagon}}$)~\cite{jiang_meron, wahl2024}. Furthermore, we add the inversion-symmetry breaking perturbation $\Delta$, which also breaks time-reversal symmetry, while keeping $\mathcal{C}_2\mathcal{T}$ symmetry intact. In real space, this term amounts to complex hoppings with Peierls phases $\phi = \pm\frac{\pi}{2}$, equivalent to a perturbation Hamiltonian
\beq{}
    H_{\Delta}  = - \frac{\Delta}{2} \sum_{\substack{\langle \langle \langle i, j \rangle \rangle \rangle_{\hexagon}}} e^{i \phi}  c^\dagger_{i} c_{j}.
\eeq
In the subsequent sections, we analyze variations in the perturbation parameter $\Delta$  and demonstrate the emergence of second-order Hall effects, which dominate the third-order quantum-geometric Hall contributions for small, i.e., perturbative, electric field strengths $\vec{E}$. The nontrivial patch Euler class, $\chi = 1$, is realized by the lower two bands of the studied kagome lattice models~\cite{Jiang2021}. In this context, we specifically set $\varepsilon_i = 0$, $t = 1$, $t' = -0.2$, $t'' = 0$, for the rest of the work, including Fig.~\ref{Fig:Model}, where we demonstrate the spectrum and geometry of the model.

In Fig.~\ref{Fig:Model}(a), we show the centrosymmetric band structure realized by the model with $\Delta = 0$. For the purposes of studying noncentrosymmetric Euler semimetals, we further introduce asymmetry by setting $\Delta=0.2$, see Fig.~\ref{Fig:Model}(b). Both models realize singular quantum geometries in the multiband geometric components involving the bands hosting the Euler class, $g^{12}_{ij}$ [Figs.~\ref{Fig:Model}(c)-\ref{Fig:Model}(e)] and $g^{23}_{ij}$ [Figs.~\ref{Fig:Model}(f)-\ref{Fig:Model}(h)], which are induced by the band degeneracies. The elements $g^{nm}_{ij}$ remain invariant under the changes of parameter $\Delta$, which by construction only modifies the band energies. Unlike the weighted quantum metric $G^n_{\alpha \beta}$, these elements are purely determined by the band geometry and topology, as captured by the momentum-space derivatives of the Bloch vectors. In constrast to the singular form of $g^{12}_{ij}$ and $g^{23}_{ij}$, we note that $g^{13}_{ij} \ll g^{12}_{ij}$ throughout the BZ. Hence, $g^{13}_{ij}$ results in negligibly small contributions to wavepacket dynamics and nonlinear transport studied in the next sections.
\begin{figure*}[t]
\centering
\includegraphics[width=\textwidth]{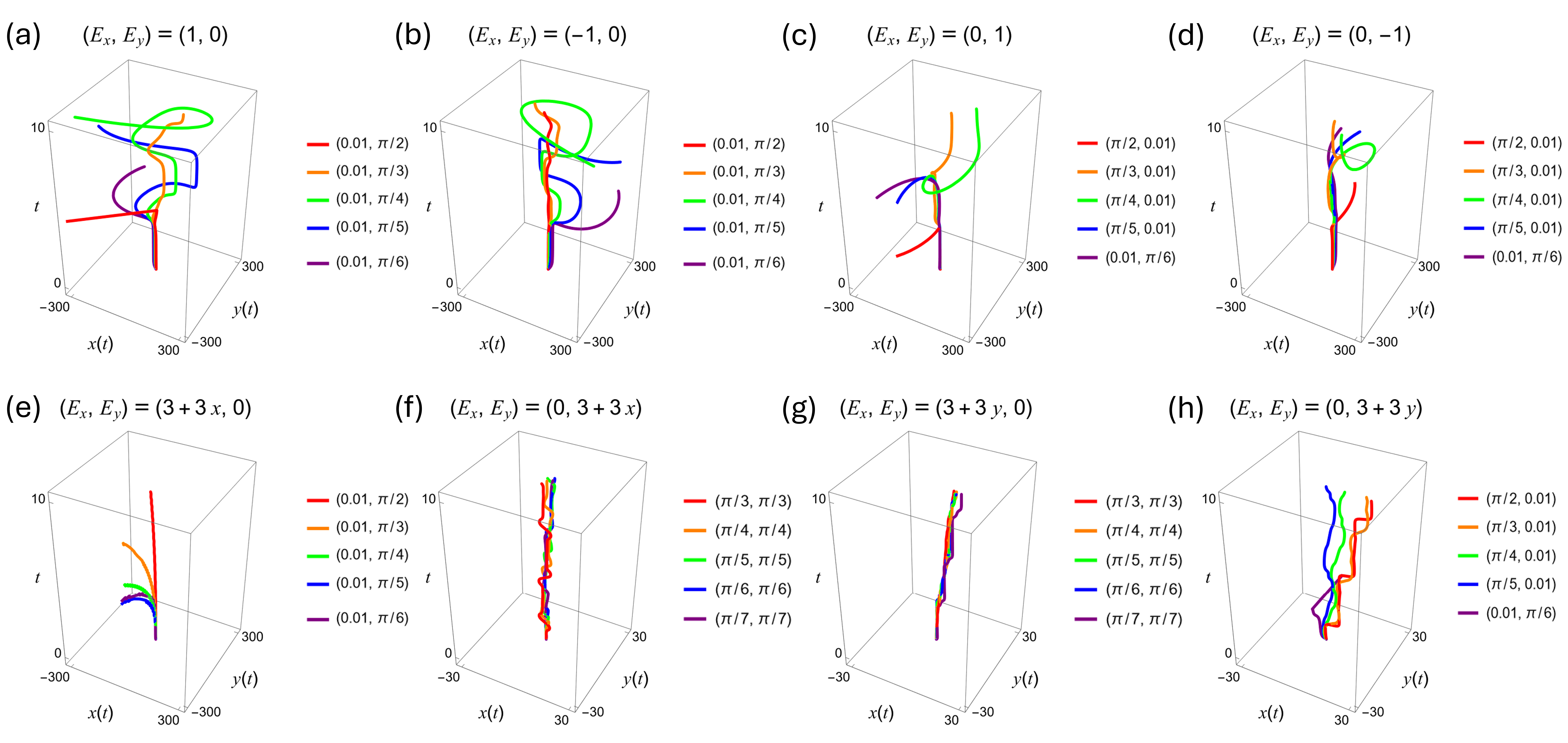}
  \caption{Anomalous wavepacket dynamics induced by the singular Euler band topology. \textbf{(a)}--\textbf{(d)} Wavepacket trajectories traced by the wavepacket center of mass $\vec{r}_c$ in two-dimensional real space plane $x-y$ as a function of time $t$ on nonlinear quadratic coupling to uniform electric field $(E^2_i)$ for different initial center-of-mass momenta $\textbf{k}_c$ with different field geometries $\textbf{E} = (E_x,E_y)$: \textbf{(a)} $(1,0)$, \textbf{(b)} $(-1,0)$, \textbf{(c)} $(0,1)$, \textbf{(d)} $(0,-1)$. \textbf{(e)}--\textbf{(h)} Wavepacket trajectories in two-dimensional real space plane $x-y$ as a function of time $t$ on coupling to constant electric field gradient $(\partial_i E_j)$ for different initial center-of-mass momenta $\textbf{k}_c$ for distinct electric fields $\textbf{E} = (E_x,E_y)$: {\textbf{(e)} $(3 + 3x,0)$}, \textbf{(f)} $(0,3+3x)$, \textbf{(g)} $(3 + 3y,0)$, \textbf{(h)} $(0,3+3y)$.}
\label{Fig:Wavepackets}
\end{figure*}

\section{Anomalous Euler wavepacket dynamics}\label{sec:IV}
We now focus on the semiclassical dynamics of individual wavepackets realizing topologically induced nontrivial quantum geometry of the non-Abelian Euler bands. We construct geodesic and geodesic deviation equations for the wavepacket dynamics, which account for the previously defined Riemannian geometry of constituent Bloch states hosting a non-Abelian topology. We consider time-dependent wavepackets of form~\cite{Gao2019a, Lapa2019}
\beq{}
    \ket{\psi_{n}(t)} = \int_{\text{BZ}} \dd^2 \kv~ a(\kv,t) e^{i \kv \cdot \hat{\vec{r}}} \ket{\tilde{u}_{n\kv}},
\eeq
in perturbed Bloch bands $\ket{\tilde{u}_{n\kv}}$ with band indices $n$, and with $\hat{\vec{r}}$ denoting the position operator (see also Appendix~\ref{app:A}). More specifically, we consider the envelope functions $|a(\kv,t)|^2 \approx \delta(\kv_c(t)-\kv)$ that are sharply centered around the center-of-mass momentum~$\kv_c$.

The geometric corrections induced by the nontrivial Euler class enter the wavepacket motion in addition to the intraband dispersion term, which is fully deduced by the band energies that can be reconstructed from ARPES experiments. While the Abelian Berry curvature vanishes in Euler bands by symmetry, the perturbative field-induced correction to the Berry curvature is nonvanishing, acts as a momentum-space gauge field strength inducing a force, and cannot be neglected in the Euler bands. The field-induced Berry curvature correction $\tilde{\vec{\Om}}_n(\textbf{E}) = \nabla_\kv \times \vec{A}'_{nn}(\textbf{E})$ is given by the gauge-invariant field-induced Berry connection correction, reading componentwise, $A^{\alpha'}_{nn} (\textbf{E}) = G^n_{\alpha \beta} E_\beta$~\cite{Gao2014}. At the same order, that is quadratic in electric field strength $(E^2)$, a band energy correction-induced weighted metric derivative emerges (see Appendix~\ref{app:A}), which, unlike the former term, yields vanishing net contributions because of the periodicity of band energies, when the response of all particles in a band is summed. Finally, the presence of nonuniform electric fields couples to the conventional quantum metric derivatives, which define momentum-space Christoffel symbols central to the geodesic equations~\cite{Smith2022}. Taken together, this yields an overall semiclassical equation of motion~\cite{Gao2014, Gao2019a, Gao2019b, Lapa2019, Smith2022},
\begin{multline}
\label{eq:derivative_semiclassical}
    (\dot{\vec{r}}_c)_\alpha = \frac{1}{\hbar} \partial_{k_\alpha} \varepsilon_{n \vec{k}} -\frac{e}{\hbar} [\vec{E} \times \tilde{\vec{\Om}}_n(\textbf{E})]_\alpha \\+ \frac{e^2}{\hbar^2} \partial_{k_\alpha} \mathcal{G}^n_{\beta\gamma} (E^2) E_\beta E_\gamma + \frac{e}{2\hbar} \partial_{k_\alpha} \mathcal{G}^n_{\beta\gamma} (\nabla) \partial_\beta E_\gamma.
\end{multline}
%
Here, we implicitly assumed the well-known coupled semiclassical equation of motion for the wavepacket center-of-mass momentum that reads
\beq{}
\label{eq:kdot}
    (\dot{\vec{k}}_c)_\alpha = \frac{1}{\hbar} F_\alpha(\textbf{r}_c) = -\frac{e}{\hbar} E_\alpha(\textbf{r}_c),
\eeq
with a force $F_\alpha(\textbf{r}_c)$ coupling to the wavepacket center-of-mass $\textbf{r}_c$, as given by the electric field $E_\alpha(\textbf{r}_c)$. In this work, we assume that the external magnetic fields are vanishing. Hence, the time-reversal symmetry, $\mathcal{T}^2 = -1$, is $not$ explicitly broken, but can be broken spontaneously. 

Furthermore, using this generalized semiclassical geodesic equation (see Appendix~\ref{app:A} for the derivation), we derive a geodesic deviation equation (Appendix~\ref{app:B}), which introduces the Riemann tensor and is retrieved in the third-order Hall response central to the next section. The momentum-space geodesic deviation equation reads
\beq{}
     \partial_{k_\delta} (\dot{\vec{r}}_c)_\gamma - \partial_{k_\gamma} (\dot{\vec{r}}_c)_\delta = \mathcal{R}^n_{\alpha \beta \gamma \delta}  (\dot{\vec{k}}_c)_\alpha (\dot{\vec{k}}_c)_\beta,
\eeq
which explicitly captures the deflection of the velocities of wavepackets with different center-of-mass momenta $\dot{\vec{k}}_c$ induced by the momentum-space Riemann curvature $\mathcal{R}^n_{\alpha \beta \gamma \delta}$. Similarly, here $(\dot{\vec{k}}_c)_\beta$ can be controlled with the electric field strengths $E_\beta$, as follows from the semiclassical equation of motion Eq.~\eqref{eq:kdot}.

In Fig.~\ref{Fig:Wavepackets}, we present the anomalous evolution of the Euler band wavepackets described by the geodesic equation induced in a model Euler Hamiltonian on the kagome lattice introduced in the previous section. We solve the coupled differential equations numerically (as detailed in Sec.~\ref{sec:III}), and identify the momentum-space horizons for the quantum metric (see Appendix~\ref{app:D}), which separate the BZ regions for center-of-mass wavepacket momenta $\vec{k}_c$ to fall, or time-evolve, into the singular quantum states at the band nodes. We further analytically support these findings using an effective $\textbf{k} \cdot \textbf{p}$ model for an Euler singularity (Appendix~\ref{app:E}). Physically, crossing the momentum-space horizon corresponds to the singular interband transfer of the wavepackets, where the initial wavepackets were constructed in a given band $n$. Manifestly, the metric derivatives governing the wavepacket evolution fingerprint the non-Abelian Berry connections on asymptotic completion to a flat (constant) metric at the $k$-space infinity within the effective continuum model, and hence allow to complete reconstruction of the patch Euler invariant. Upon combining the responses of individual wavepackets realized in the bands, the net bulk transport responses can also be obtained (see Appendix~\ref{app:G}). We address such anomalous responses in the next section.

As mentioned before, centrally to this work, the retrieved anomalous wavepacket dynamics allows one to infer the Euler class. To reconstruct the Euler invariant from the wavepacket dynamics, the following steps are undertaken. First, the trajectory $\vec{r}_c = (x(t), y(t))$ of a wavepacket forced with field $\vec{E} = (E_x, E_y)$ within a two-dimensional plane is retrieved as a function of time, such that the wavepacket velocity $\dot{\vec{r}}_c$ is known. Second, the derivatives of weighted quantum metrics $\partial_{k_\gamma} \mathcal{G}^n_{\alpha \beta}$ need to be obtained as a function of the center-of-mass momentum $\vec{k}_c$, by subtracting the group velocity contributions $\dot{\vec{r}}_c|_{\vec{E}=0} = \frac{1}{\hbar} \nabla_\kv \varepsilon_{n\kv}$, based on the known band energies $\varepsilon_{n\kv}$. Third, the multiband metric components $g^{nm}_{\alpha \beta}$ need to be extracted from $\partial_{k_\gamma} \mathcal{G}^n_{\alpha \beta}(\nabla)$ and $\partial_{k_\gamma}  \mathcal{G}^n_{\alpha \beta}(E^2)$ derivatives, following the aforementioned asymptotic completion argument. The deduced metric components $g^{n(n+1)}_{\alpha \alpha} = |a_\alpha|^2$ fully determine the Euler connection $a = a_\alpha \text{d} k_\alpha$ in real gauge. The invariant can then be calculated using Eq.~\eqref{eq:eq1}, similarly to the strategy of Ref.~\cite{jankowski2023optical}, where instead of being deduced from wavepacket dynamics, the $g^{n(n+1)}_{\alpha \alpha}$ components are probed optically. For further mathematical details on the reconstruction of the Euler invariant with electric fields, beyond the discussed key steps, see Appendix~\ref{app:F}.

We now show how the reconstruction scheme can be implemented in the context of Fig.~\ref{Fig:Wavepackets}. In Figs.~\ref{Fig:Wavepackets}(a)-\ref{Fig:Wavepackets}(d), the time-dependent evolution of the wavepacket center-of-mass position $\vec{r}_c$ for different center-of-mass momenta $\vec{k}_c$, under the constant electric fields $\vec{E}$ pointing individually in positive and negative $x$ and $y$ directions, are presented. In Figs.~\ref{Fig:Wavepackets}(e)-\ref{Fig:Wavepackets}(h), an addition of electric field gradients targeting $\partial_{k_\gamma} \mathcal{G}^n_{\alpha \beta}(\nabla)$ coefficients is included. We observe that in Figs.~\ref{Fig:Wavepackets}(a)-\ref{Fig:Wavepackets}(d), the $\partial_{k_\gamma}  \mathcal{G}^n_{\alpha \alpha}(E^2)$ components drive increasing deviations for both $\gamma=x,y$, as $\vec{k}_c$ approaches the momentum of the Euler node. The increasing values of $\partial_{k_\gamma}  \mathcal{G}^n_{\alpha \alpha}(E^2)$, which reflect the increasing changes of $g^{nm}_{\alpha \alpha}$ presented in Fig.~\ref{Fig:Model}, result in the distinct spinning motions as $\vec{k}_c \rightarrow 0$, depicted in different colors, from red to purple. In Figs.~\ref{Fig:Wavepackets}(e), and \ref{Fig:Wavepackets}(h), we analogously observe that the electric field gradients $\partial_\alpha E_\alpha$ cause anomalous spinning wavepacket motions by coupling to  $\partial_{k_\gamma} \mathcal{G}^n_{\alpha \alpha}(\nabla)$ according to Eq.~\eqref{eq:derivative_semiclassical}. Similarly to the previous case, the high values of $\partial_{k_\gamma} \mathcal{G}^n_{\alpha \alpha}(\nabla)$ on approaching $\vec{k}_c \rightarrow 0$ reflect the singular features of $g^{nm}_{\alpha \alpha}$, as shown in Figs.~\ref{Fig:Model}(c) and \ref{Fig:Model}(d). In contrast, Figs.~\ref{Fig:Wavepackets}(f), and \ref{Fig:Wavepackets}(g) show reduced anomalous deviations with respect to the group velocity term $\dot{\vec{r}}_c|_{\vec{E}=0} = \frac{1}{\hbar} \nabla_\kv \varepsilon_{n\kv}$, showing that the coupling of $\partial_x E_y$ and $\partial_y E_x$ to $\partial_{k_\gamma} \mathcal{G}_{xy}(\nabla)$, cancels with the simultaneously present contributions of $\partial_{k_\gamma} \mathcal{G}^n_{xy}(E^2)$, consistently with the opposite sign of $g^{nm}_{xy}$ components shown in Fig.~\ref{Fig:Model}, and consistently with the continuum model expressions of Appendix~\ref{app:E}. As discussed within the introduced step-by-step scheme, the extracted  $g^{nm}_{\alpha \alpha}$ components allow reconstruction of the Euler invariant, analogously to the optical protocol of Ref.~\cite{jankowski2023optical}.

\section{Bulk anomalous currents from non-Abelian bands}\label{sec:V}
Further to the anomalous dynamics of individual wavepackets governed by the quantum-geometric geodesic equation, we study bulk nonlinear responses induced by the quantum geometry of the Euler phases at different voltage-controllable dopings. With $\mathcal{C}_2 \mathcal{T}$ symmetry necessary for the realization of the fermionic Euler invariant, two scenarios emerge (i) when $(\mathcal{C}_2\mathcal{T})^2 = 1$, with $(\mathcal{C}_2)^2 = -1$ and $\mathcal{T}^2 = -1$, or (ii) $(\mathcal{C}_2\mathcal{T})^2 = 1$, with $(\mathcal{C}_2)^2 \neq -1$ and $\mathcal{T}^2 \neq -1$, i.e., both \text{individual} crystalline and time-reversal symmetries are implicitly broken in the free Hamiltonian. In the latter case, second-order responses constitute a dominant contribution within the anomalous transport. Under inversion symmetry, which is equivalent to the $\mathcal{C}_2$ symmetry in two spatial dimensions, the second-order currents vanish, and the third-order Hall response becomes dominant. In the noncentrosymmetric case, the quantum-geometric second-order nonlinear current densities can emerge. For noncentrosymmetric cases, we consider the same models with inversion-symmetry-breaking perturbations of strength $\Delta$, which yield a modified Hamiltonian  $H(\kv) \rightarrow H(\kv) + \Delta \sin k_x \mathsf{1}$, with $\mathsf{1}$ the identity matrix. Under the nonuniform electric fields, in inversion-free systems, we evaluate geometric currents, $j_\gamma(\nabla) = \sigma_{\gamma, \alpha \beta} (\nabla) \partial_\alpha E_\beta$, following Ref.~\cite{Lapa2019}, with the corresponding dispersive conductivity tensor contributions,
\beq{}
    \sigma_{\gamma, \alpha \beta} (\nabla) = -\frac{e^2}{2\hbar} \int\limits_\kv \sum_n f_{n\kv}(\mu)~ \partial_{k_\gamma} \mathcal{G}^n_{\alpha \beta} (\nabla), 
\eeq
where $\int_\kv \equiv \intBZ \frac{\dd^2 \textbf{k}}{(2\pi)^2}$ denotes the momentum-space integral over the BZ. Here, $f_{n\kv}(\mu) = [\text{exp}(\frac{\varepsilon_{n\kv}-\mu}{T}) + 1]^{-1}$ is a filling factor given by the chemical potential ($\mu$) within the equilibrium Fermi-Dirac distribution.
For second order in electric field, we acquire geometric currents $j_\gamma(E^{2}) = \sigma_{\gamma, \alpha \beta} (E^{2}) E_\alpha E_\beta$. Here, we focus on the quantum-metric dipole-induced Hall response contributed by the gradients of $\mathcal{G}^n_{\alpha \beta}(E^2)$~\cite{Gao2014, Wang_2023, Kaplan2024},
\beq{}
\begin{split}
    \sigma_{\gamma, \alpha \beta}(E^{2}) = -\frac{e^3}{\hbar^2} \int\limits_\kv  \sum_n f_{n\kv}(\mu) \Big[ \partial_{k_\gamma} \mathcal{G}^n_{\alpha \beta} -\frac{1}{2} \partial_{k_{(\alpha}} \mathcal{G}^n_{\beta)\gamma} \Big], 
\end{split}
\eeq
where $(\ldots)$ denotes an unnormalized symmetrization with respect to the spatial indices.
Effectively, one obtains such nonlinear Hall response from the BZ integration of the individual wavepacket contributions. In noncentrosymmetric Euler semimetals, such response arises from the topologically-induced net quantum-metric dipole present in the integrand, as induced by the singular Euler node geometry (see Appendix~\ref{app:H}).

In the other case, when inversion and time-reversal symmetry are present, the second order responses and the responses to nonuniform electric field vanish because of the transformation of the forcing vectors and corresponding conductivities under the inversion symmetry. In that case, the dominant quantum-geometric current contribution of interest emerges from the \textit{third-order} responses~\cite{Lai2021, Mi2023, Mandal2024, Li2024}. More concretely, the quantum-geometric third-order contribution to currents, ${j_\delta(E^{3}) = \sigma_{\delta, \alpha \beta \gamma} (E^{3}) E_\alpha E_\beta E_\gamma}$, amounts to the third-order quantum-metric quadrupole-induced Hall conductivity that reads~\cite{Mandal2024},
\begin{align}
    \sigma_{\delta, \alpha \beta \gamma}(E^3) &=-\frac{2e^4 \tau}{3\hbar^3} \int\limits_\kv \sum^{}_{n} f_{n \kv}(\mu) \\& \times\Big[\partial_{k_\beta} \partial_{k_{(\alpha}} \mathcal{G}^{n}_{\gamma \delta)} - \partial_{k_{(\gamma}} \partial_{k_{\delta}} \mathcal{G}^{n}_{\alpha)\beta}\Big]\nonumber.
\end{align}
Here, $\tau$ is the scattering time, as the third-order quantum-geometric Hall conductivity contributions in the $\mathcal{C}_2\mathcal{T}$-symmetric systems with inversion symmetry are only extrinsic, unlike the second-order quantum-geometric Hall conductivity contributions~\cite{Mandal2024}. The third-order conductivity acquires contributions from the generalized Riemann curvature tensor $\mathcal{R}^n_{\alpha \beta \gamma \delta}$. For more details, see Appendices~\ref{app:E} and~\ref{app:G}. Centrally to this work, the introduced geometric Hall conductivity contributions allow access to the Euler class $\chi$ of a semimetallic node.

The Euler invariant $\chi$ can be inferred from the nonlinear conductivity measurements in the following steps. First, one identifies the presence of the relevant singular geometric nonlinear Hall conductivity contribution, where from a phenomenological perspective, each geometric contribution can be characterized by a unique scaling with scattering times $\tau$~\cite{Mandal2024}. Secondly, we note the strength of the measured singular contributions depends on the Euler charge $\chi$, see Appendix~\ref{app:H} for the term-by-term analytical identifications within the continuum models. Hence, by changing the doping with the chemical potential $\mu$, one can observe the change of the corresponding geometric Hall conductivity contribution, depending on charge $\chi$, which is reflected by an integrated singular momentum-space scaling provided by the quantum metric $g^{n(n+1)}_{\alpha \alpha}$ (see Appendix~\ref{app:E}). The singular metric determines the conductivity for the proximity of the Fermi level around the Euler node (see Appendix~\ref{app:H}).

In Fig.~\ref{Fig:Bulk}, we show the corresponding results in the zero-temperature limit for different chemical potentials $\mu$ in both inversion-symmetric and inversion-symmetry-free kagome Euler Hamiltonians. We observe that changing the doping allows one to explicitly infer the geometry of the non-Abelian bands over BZ, which ultimately allows to deduce the Euler invariant to be $\chi=1$ in the model Hamiltonians. Finally, we retrieve the chemical potential-dependent scaling of these bulk third-order and second-order currents in the effective continuum Euler Hamiltonians in Appendix~\ref{app:H}, which analytically supports the numerical results in realistic lattice-regularized Hamiltonians. 

The individual bulk conductivities of Fig.~\ref{Fig:Bulk}, which reflect the Euler class $\chi=1$, can be understood as follows. Figure~\ref{Fig:Bulk}(a) shows the bulk scaling of conductivity in response to electric field gradients~\cite{Lapa2019}. The response function is nonanalytic around the energy of the Euler node $(\mu = -0.8)$, which demonstrates the anomalous enhancement in derivatives $\partial_{k_\gamma} \mathcal{G}_{\alpha \alpha}^n(\nabla)$ induced by the singular $g^{n(n+1)}_{\alpha \alpha}$. Notably, the response requires breaking inversion symmetry, and its magnitude heavily depends on the degree of noncentrosymmetry, as captured by the parameter $\Delta$. Analogously to the wavepacket dynamics, the $\sigma_{\gamma,\alpha\alpha}$ responses have opposite signs to $\sigma_{\gamma,xy}$, given the opposite sign of the quantum metric components, as shown in Fig.~\ref{Fig:Model}(e). In Fig.~\ref{Fig:Bulk}(b), we show the second order Hall responses, which also depend on the size of $\Delta$, by symmetry. The scaling of nonlinear Hall conductivity reflects $\partial_{k_\gamma} \mathcal{G}_{\alpha\beta}^n(E^2)$ derivatives, which are more singular than $\partial_{k_\gamma} \mathcal{G}_{\alpha\beta}^n(\nabla)$ due to introduced weights $c_{nm} = \hbar/(\varepsilon_{n \kv} - \varepsilon_{m \kv})$ that diverge in the limit of $\kv$ approaching the Euler node.  Finally, Fig.~\ref{Fig:Bulk}(c) demonstrates the scaling behavior of the geometric third-order nonlinear Hall conductivity term, which reflects the patch Euler class $\chi$ even in the centrosymmetric model (see Appendix~\ref{app:H}). Similarly, the divergence is induced by the singular second derivatives $\partial_{k_\delta} \partial_{ k_\gamma} \mathcal{G}_{\alpha\beta}^n(E^2)$ which, beyond confirming the presence of $g^{n(n+1)}_{\alpha \alpha}$ metric singularity, reflect the singular Riemann curvature $\mathcal{R}^n_{\alpha \beta \gamma \delta}$ and the divergent Kretschmann invariant that can be associated with the Euler node, see Appendix~\ref{app:E}. Further quantitative details on the exact form of the geometric conductivities against the chemical potential $\mu$, as induced by the Euler node within an effective low-energy continuum Hamiltonian picture, are provided in Appendix~\ref{app:H}.
\begin{figure}[t!]
\centering
\includegraphics[width=0.92\columnwidth]{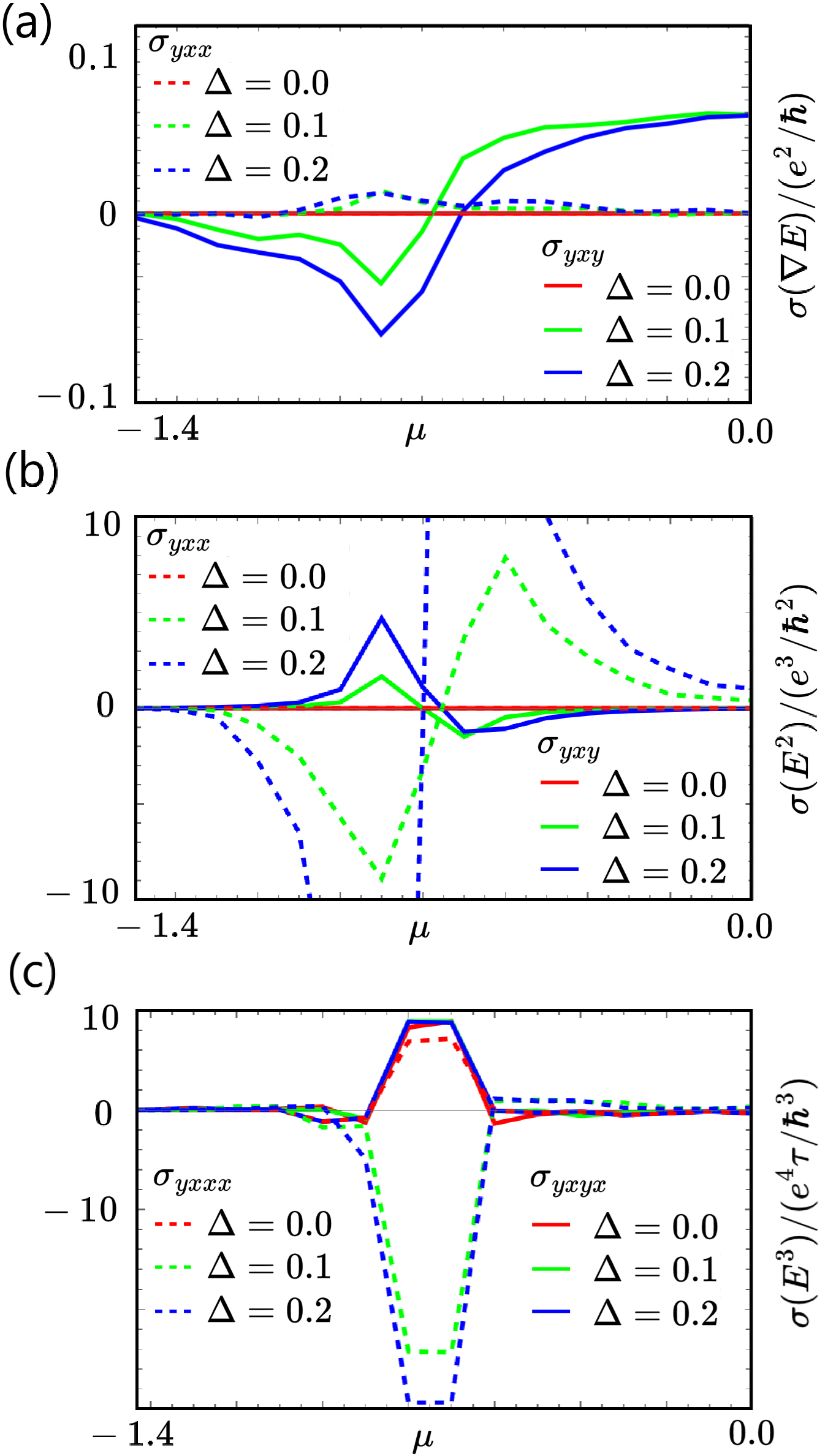}
  \caption{Bulk anomalous Hall responses induced by the nontrivial Euler class associated with a topological band node. The band node appears at the chemical potential $\mu = -0.8$ within the kagome lattice model. We retrieve the response to the nonuniform electric fields \textbf{(a)}, and uniform electric fields \textbf{(b)}--\textbf{(c)} at nonlinear (i.e., second/third) orders. We compute the anomalous conductivities at different chemical potentials $\mu$ which fix the band occupations in the Euler Hamiltonian on a kagome lattice for different strengths ($\Delta$) of inversion-symmetry breaking terms. \textbf{(a)}--\textbf{(b)} The currents in response to electric field gradients $\partial_\alpha E_\beta$ and geometric Hall currents quadratic in $E_\alpha$ emerge when $\Delta \neq 0$. \textbf{(c)} On the contrary, the third-order Hall response arises even when the inversion symmetry is not broken.}
\label{Fig:Bulk}
\end{figure}

\section{Conductance and Fermi surface topology} Finally, following the insights of Ref.~\cite{Kane_2022}, we consider nonlinear ballistic conductance signatures of the Euler phases, emerging from the Fermi surface topology, on doping away from the Euler nodes. The Euler characteristic $\chi_F$ specifically captures the topology of the Fermi surface defined by the energies of the occupied single-particle states over the momentum space. On the other hand, the Euler class $\chi$, captures the topology encoded in the geometric winding of the Bloch vectors, as viewed in the real gauge. Furthermore, $\chi$ fingerprints the associated singular band geometry through its relations to geometric connections and curvatures~\cite{bouhon2023quantumgeometry, jankowski2023optical}. As pointed out in the context of graphene, the presence of band nodes allows to control the Euler characteristic $\chi_F$ of electron and hole Fermi surfaces. A single band crossing allows to introduce $\chi_F = \text{sgn}(E_F-E_0)$ on doping away from the band node energy $E_0$. Furthermore, the intrinsic nonlinear conductance $\alpha_i(\om_1, \om_2)$ in response to electric potential perturbations $V_1(\om_1)$, $V_2(\om_2)$ with frequencies $\om_1, \om_2$ applied across a triple contact was shown to be quantized by the Euler characteristic $\chi_F$ of the Fermi surface~\cite{Kane_2022}. We note that contrary to graphene, the presence of nontrivial Euler class $\chi$ allows for $2|\chi|$ topologically-protected linear nodes of same chirality, which formally is a consequence of Poincar\'e-Hopf index theorem~\cite{bouhon2020geometric}. In the Euler phases with topologically-protected underdoped linear nodes, hence with $\chi_F = 2|\chi|$, the intrinsic nonlinear conductance $\alpha_i(\om_1, \om_2)$ therefore reads~\cite{Kane_2022},
\beq{}
    \alpha_i(\om_1, \om_2) =\frac{2e^3}{\hbar^2} |\chi| \frac{1}{(\om_1-i 0^+)(\om_2-i 0^+)}.
\eeq
We conclude by noting that unlike in the case of graphene, where nodes of opposite chiralities can be pair-annihilated, a similar attempt to mutually annihilate linear nodes in Euler phases results in merging these into a quadratic node, as long as the $\mathcal{C}_2\mathcal{T}$ symmetry is preserved. Hence, unlike in graphene, the quantized ballistic conductance of underdoped Euler semimetals can result in the change of parity of associated $\chi_F$, as two nodes result in one, rather than in none~\cite{BJY_nielsen, bouhon2020geometric}. This observation directly reflects the circumvention of Nielsen-Ninomiya fermion doubling theorem admitted by the systems with nontrivial Euler class~\cite{BJY_nielsen, bouhon2019nonabelian}. As we predict here, this should in principle be directly observable under the triple-contact conductance measurements of Euler phases, through the characteristic parity flip of the quantized nonlinear response. Importantly, such parity-flipping through the merging, or splitting, of non-Abelian band nodes~\cite{bouhon2019nonabelian, Jiang2021} could be achieved by applying tensile strains under the experimental conditions applicable to real materials.

\section{Discussion}\label{sec:VI}
In this work, we considered anomalous bulk and wavepacket responses induced by the geometry and topology of Euler bands. We anticipate our results to be of relevance to kagome materials~\cite{Kang2024} and honeycomb multilayers~\cite{mondal2024}, in the cases in which the interactions do not spontaneously break $\mathcal{C}_2 \mathcal{T}$ symmetry. We note that our findings allow to corroborate the nodal topology and quantum geometry present in layered two-dimensional systems, e.g., in Bernal-stacked bilayer graphene realizing optical conductivity consistent with the presence of patch Euler class $\chi = 1$~\cite{Nicol2008, jankowski2023optical} around the quadratic band touchings. It should be stressed that if $\mathcal{T}$~symmetry is broken, breaking the $\mathcal{C}_2 \mathcal{T}$ symmetry in Euler bands can induce the formation of Chern bands~\cite{bouhon2022multigap}, which were experimentally and theoretically studied in a range of kagome systems~\cite{Sun2011, Xu2015, Guo_2022}. Alternatively, breaking $\mathcal{C}_2$~symmetry, but preserving $\mathcal{T}$~symmetry, could yield second-order nonlinear Berry curvature dipole Hall responses~\cite{Sodemann2015}, see Appendix~\ref{app:G}.

While the stability of the presented unconventional transport signatures of Euler semimetals against disorder~\cite{jankowskiPRB2024disorder} and interactions naturally present in realistic materials is left for future study, we expect the geometric enhancement of the bulk responses to be a robust feature, as long as the quantization of the Euler invariant is present. Another important aspect arises from the fact that the Euler topology can be realized in a pair of fully degenerate bands over BZ, where it is equivalent within a unitary transformation to a pair of opposite Chern bands $C = \pm \chi$ with the opposite Wilson loop windings protected by the $\mathcal{C}_2 \mathcal{T}$ symmetry~\cite{wahl2024}. In such case, no bulk Hall responses are expected, as the opposite Chern sectors ensure that the Hall responses cancel at any order in electric field. However, the fully degenerate limit over BZ remains singular and highly-tuned, i.e., the electronic structures of real materials break such model degeneracies, as known from the context of magic-angle twisted bilayer graphene (MATBG)~\cite{Potwisted}, for example. As supported by the Poincar\'e-Hopf theorem, in the nondegenerate limit~\cite{bouhon2020geometric}, the Euler nodes need to remain as long as the symmetry is preserved. In such cases, any filling controlled by external voltage allows one to access the transport features underpinned by the Euler-invariant-induced singular quantum geometry, as shown in this work. As such, we expect that even minimal doping of nondegenerate Euler bands in real materials should allow access to the invariant, as demonstrated explicitly in the previous sections.

\section{Conclusions}\label{sec:VII}
We unravel nonlinear transport signatures induced by the non-Abelian topologies of Euler bands and associated Riemannian quantum geometry. We obtain the manifestations of Euler topology via geodesic equations and geodesic deviations of individual wavepackets. Furthermore, we find nonlinear bulk Hall currents at second and third orders in electric field, as induced by the quantum geometry of Euler semimetals in the presence and absence of individual inversion and time-reversal symmetries. We anticipate that our results could serve as smoking-gun probes of exotic non-Abelian band topology in two-dimensional, e.g., kagome or honeycomb multilayer, systems.

\begin{acknowledgements}
    The authors cordially thank Adrien Bouhon, F. Nur \"Unal, Giandomenico Palumbo, and Gaurav Chaudhary, for numerous discussions on non-Abelian band topologies and quantum geometry. W.J.J.~acknowledges funding from the Rod Smallwood Studentship at Trinity College, Cambridge. R.-J.S. acknowledges funding from an EPSRC ERC Underwrite Grant  EP/X025829/1, and a Royal Society Exchange Grant IES/R1/221060 as well as Trinity College, Cambridge.
\end{acknowledgements}

\appendix


\section{Derivation of the momentum-space geodesic equation}\label{app:A}

We derive the geodesic equation on introducing perturbative field-induced energy and eigenstate corrections. Here, we consider the following perturbation Hamiltonian:
\beq{}
    \Delta \hat{H} = \Delta \hat{H}^{(1)}_E + \Delta \hat{H}^{(2)}_{\partial E} = e \hat{\vec{r}} \cdot \vec{E} + e \vec{\hat{r}} \cdot (\vec{\hat{r}} \cdot \nabla) \vec{E},
\eeq
where we assume that the electric-field gradient perturbation that arises from the nonuniformity of electric field is weaker than the electric field perturbation itself. In this work, we consider all the coupling to electric fields manifestly in length gauge. Within the perturbative expansion, we have: $\varepsilon_{n \vec{k}} \rightarrow \tilde{\varepsilon}_{n \vec{k}} = \varepsilon^{(0)}_{n \vec{k}} + \varepsilon^{(1)}_{n \vec{k}} + \varepsilon^{(2)}_{n \vec{k}} + \ldots$, with $\varepsilon^{(0)}_{n \vec{k}} = \varepsilon_{n \vec{k}}$, the unperturbed energy. and the perturbative corrections to the eigenstates $\ket{\psi_{n\kv}} = e^{i \kv \cdot \hat{\vec{r}}} \ket{u_{n\kv}}$: $\ket{u_{n\kv}} \rightarrow \ket{\tilde{u}_{n\kv}} = \ket{u^{(0)}_{n\kv}} + \ket{u^{(1)}_{n\kv}} + \ket{u^{(2)}_{n\kv}} + \ldots$, both of which modify the wavepacket group velocity $\dot{\vec{r}}_c$.  The first-order correction to Bloch eigenstates reads,
%
\begin{multline}    
    \ket{u^{(1)}_{n\kv}} = \frac{\bra{\psi_{m\kv}} \Delta \hat{H} \ket{\psi_{n\kv}}}{\varepsilon^{(0)}_{n\kv}-\varepsilon^{(0)}_{m\kv}} \ket{u^{(0)}_{m\kv}}  \\ \approx  \frac{e \vec{E} \cdot \vec{A}_{nm}}{\varepsilon^{(0)}_{n\kv}-\varepsilon^{(0)}_{m\kv}} \ket{u^{(0)}_{m\kv}} \equiv e E_\alpha M^\alpha_{nm} \ket{u^{(0)}_{m\kv}},
\end{multline}
%
where $\vec{A}_{nm} \equiv i \bra{u_{n \kv}} \ket{\nabla_\kv u_{m \kv}}$ is the non-Abelian Berry connection, and $M^\alpha_{nm} \equiv \frac{A^\alpha_{nm}}{\varepsilon^{(0)}_{n\kv}-\varepsilon^{(0)}_{m\kv}} (1-\delta_{nm})$.

Following the notation introduced in the main text, the wavepacket state reads
\beq{}
    \ket{\psi_{n}(t)} = \int_\kv e^{i \kv \cdot \vec{r}} \Big( a_n(\kv, t)  \ket{u^{(0)}_{n\kv}}  + \sum_{m \neq n} a_m(\kv, t) \ket{u^{(0)}_{m\kv}} \Big),
\eeq
with $|a_m(\kv, t)| \approx e E_\alpha |M^\alpha_{nm}|$. The center-of-mass position of the wavepacket reads $\vec{r}_c(t) = \bra{\psi_{n}(t)} \hat{\vec{r}}\ket{\psi_{n}(t)}$, whereas the center-of-mass momentum reads, $\vec{k}_c(t) = \bra{\psi_{n}(t)} \hat{\vec{p}}\ket{\psi_{n}(t)}$, with momentum $\hat{\vec{p}} = -i\nabla_\vec{r}$. The energy of the wavepacket reads $\tilde{\varepsilon}_{n\kv}= \bra{\psi_{n}(t)} \hat{H} \ket{\psi_{n}(t)}$, with $\hat{H} = \hat{H}_0 + \Delta \hat{H}$, and for brevity, we write $\kv = \kv_c$ for the center-of-mass momentum.

The first-order energy correction reads,
%
\begin{multline}   
     \varepsilon^{(1)}_{n \vec{k}} =  \bra{\psi_{n} (t)} \Delta H \ket{\psi_{n} (t)} = e E_\alpha \cdot (\vec{r}_c)_\alpha  \\ + e \partial_{(\alpha} E_{\beta)} \bra{\psi_{n}(t)} (\hat{r})_\alpha (\hat{r})_\beta \ket{\psi_{n}(t)}.
\end{multline}
%
where we introduced an electric-field gradient $\partial_{(\alpha} E_{\beta)}$ symmetrized in spatial indices $\alpha,\beta$. Following Ref.~\cite{Lapa2019}, we utilize the relation between the spread of the wavepacket and the quantum metric; $\bra{\psi_{n}(t)} \hat{r}_\alpha \hat{r}_\beta \ket{\psi_{n}(t)} \approx (\vec{r}_c)_\alpha (\vec{r}_c)_\beta + g^n_{\alpha \beta}(\kv_c)$. Notably, here, only the metric term $g^n_{\alpha \beta}(\kv_c)$ enters the group velocity for the energy correction $\varepsilon^{(1)}_{n \vec{k}}$.
The second-order energy correction, consistently with the time-independent perturbation theory, reads
\beq{}
     \varepsilon^{(2)}_{n \vec{k}} =  \sum_{m \neq n} \frac{|\bra{\psi_{n\kv}} \Delta H \ket{\psi_{m\kv}}|^2}{\varepsilon^{(0)}_{n\kv}-\varepsilon^{(0)}_{m\kv}} = - e^2 E_\alpha G^n_{\alpha \beta} E_\beta,
\eeq
where $G^n_{\alpha \beta} = \hbar~\text{Re} \sum_{m \neq n} \frac{A^\alpha_{nm} A^\beta_{mn}}{\varepsilon^{(0)}_{n\kv}-\varepsilon^{(0)}_{m\kv}} = \hbar~\sum_{m \neq n} \frac{g^{nm}_{\alpha \beta}}{\varepsilon^{(0)}_{n\kv}-\varepsilon^{(0)}_{m\kv}}$, the weighted quantum metric defined in the main text. Here, we recognize that $\bra{\psi_{m\kv}} \textbf{r} \ket{\psi_{n\kv}} = i \bra{u_{m\kv}} \ket{\nabla_\kv u_{n\kv}} = A_{mn} (\kv)$ and employ the weighted quantum geometric tensor (QGT) $\mathcal{Q}_{\alpha \beta}^{n}(\kv) = \sum_m c_{nm} A^\alpha_{nm} A^\beta_{mn}$, with weights $c_{nm} = \hbar/(\varepsilon_{n \kv} - \varepsilon_{m \kv})$. We now combine the energy corrections and utilize the notation from the main text $\mathcal{G}_{\alpha \beta}^{n}(E^2) \equiv G^n_{\alpha \beta}$, and $\mathcal{G}_{\alpha \beta}^{n}(\nabla) \equiv g^n_{\alpha \beta}$. On differentiating with respect to the center-of-mass momentum $\vec{k}_c$, we have,
%
\begin{multline}   
     \partial_{k_\gamma} \tilde{\varepsilon}_{n \vec{k}} \approx  \partial_{k_\gamma} \varepsilon_{n \vec{k}} + \frac{e^2}{\hbar} \partial_{k_\gamma} \mathcal{G}^{n}_{\alpha \beta}(E^2) E_\alpha E_\beta \\ + \frac{e}{2} \partial_{k_\gamma} \mathcal{G}^{n}(\nabla)_{\alpha \beta}(E^2) \partial_{\alpha} E_{\beta}. 
\end{multline}
%
Notably, the symmetry in electric fields picks the symmetric parts of the QGTs, i.e., the weighted quantum metrics $\mathcal{G}_{\alpha \beta}^{n}(\kv) \equiv \text{Re}~\mathcal{Q}_{\alpha \beta}^{n}(\kv)$. Having deduced the wavepacket energy corrections, we now deduce the wavepacket velocity satisfying a semiclassical equation of motion. Using time-dependent Schr\"odinger equation, and the definition of the center-of-mass position $\dot{\vec{r}}_c$, the wavepacket velocity reads $(\dot{\vec{r}}_c)_\alpha = \bra{\psi_n(t)} \frac{i}{\hbar} [H, \hat{r}_\alpha] \ket{\psi_n(t)} \approx  \frac{1}{\hbar} \bra{\tilde{u}_{n\kv}} \partial_{k_\alpha} H \ket{\tilde{u}_{n\kv}}$. Apart from the energy correction, which is symmetric in spatial indices, antisymmetric corrections associated with the perturbed Berry curvature can arise. Here, the Berry curvature enters through interband terms,
\begin{multline}
    \bra{u^{(0)}_{n\kv}} \partial_{k_\alpha} H \ket{u^{(1)}_{n\kv}} + \text{c.c.}  
\\ =  e E_\beta \sum_m A^\beta_{nm} \frac{\bra{u^{(0)}_{n\kv}} \partial_{k_\alpha} H \ket{u^{(0)}_{m\kv}}}{\varepsilon^{(0)}_{n\kv}-\varepsilon^{(0)}_{m\kv}} + \text{c.c.} \\ = -i e E_\beta \sum_mA^\beta_{nm} A^\alpha_{mn} + \text{c.c.}  = -e E_\beta \Om^n_{\beta \alpha}, 
\end{multline}
with Berry curvature $\Om^n_{ \alpha \beta} = -2 \text{Im} \sum_m A^\alpha_{nm} A^\beta_{mn}$, which defines the pseudovector: $(\vec{\Om}_n)_\gamma = (\nabla_\kv \times \vec{A}_{nn})_\gamma = \frac{1}{2} \epsilon_{\alpha \beta \gamma} \Om^n_{\alpha \beta}$. Similarly to the energies, the Berry curvature becomes perturbed at nonlinear order $\vec{\Om_n} \rightarrow \vec{\Om_n} + \vec{\Om'_n}$, with $\vec{\Om'_n} \equiv \nabla_\kv \times \vec{A}_{nn}'$, the field-induced Berry curvature correction arising from the field-induced correction to the Berry connection $\vec{A}_{nn}' = i \bra{u^{(0)}_{n\kv}}\ket{\nabla_\kv  u^{(1)}_{n\kv}} + \text{c.c.}$~\cite{Gao2014}. On inserting $\ket{u^{(1)}_{n\kv}}$: $(\vec{A}_{nn}')_\alpha = G^n_{\alpha \beta} E_\beta$~\cite{Gao2014}, which in a matrix form can be written as $\vec{\Om'_n}= \nabla_\kv \times (\underline{\underline{G^n}}\vec{E})$, where we use a double underline to denote considered matrix quantity in its matrix form. In particular, in an Euler Hamiltonian, $\vec{\tilde{\Om}_n}= \vec{\Om'_n}$, as $\vec{\Om_n}$ vanishes by $\mathcal{C}_2\mathcal{T}$  symmetry.

On combining with the definition of the center-of-mass velocity, we now arrive at,
%
\begin{multline}    
    \label{eq:geodesic1}
    (\dot{\vec{r}}_c)_\gamma = \frac{1}{\hbar} \partial_{k_\gamma} \tilde{\varepsilon}_{n \vec{k}} - \frac{e}{\hbar} [\vec{E} \times \tilde{\vec{\Om}}_n]_\gamma = \frac{1}{\hbar} \partial_{k_\gamma} \varepsilon_{n \vec{k}} - \frac{e}{\hbar} [\vec{E} \times \tilde{\vec{\Om}}_n]_\gamma \\ + \frac{e^2}{\hbar^2} \partial_{k_\gamma} \mathcal{G}^{n}_{\alpha \beta}(E^2) E_\alpha E_\beta + \frac{e}{2 \hbar} \partial_{k_\gamma} \mathcal{G}^{n}_{\alpha \beta}(\nabla) \partial_{\alpha} E_{\beta},
\end{multline}
%
which is the final result governing the semiclassical wavepacket equation of motion, implicitly retrieving the geodesic equation in momentum space. Namely, on replacing $\dot{\vec{k}}_c = - \frac{e}{\hbar} \vec{E}$, we arrive at an electric field gradient-forced form of a ``momentum-space geodesic equation"~\cite{Smith2022}, 
%
\begin{multline}
    \label{eq:geodesic2}
    (\dot{\vec{r}}_c)_\gamma = \frac{1}{\hbar} \partial_{k_\gamma} \varepsilon_{n \vec{k}} - \big[\dot{\vec{k}}_c \times \big(\nabla_\kv \times (\underline{\underline{\mathcal{G}}}^n(E^2) \dot{\vec{k}}_c)\big)\big]_\gamma \\+ \partial_{k_\gamma} \mathcal{G}^{n}_{\alpha \beta}(E^2) (\dot{\vec{k}}_c)_\alpha (\dot{\vec{k}}_c)_\beta + \frac{e}{2 \hbar} \partial_{k_\gamma} \mathcal{G}^{n}(\nabla)_{\alpha \beta}(E^2) \partial_{\alpha} E_{\beta},
\end{multline}
%
which completes the derivation central to this section. 

\section{Derivation of the momentum-space geodesic deviation equation}\label{app:B}

Here, we derive the momentum-space geodesic deviation equation in the uniform electric fields. We furthermore show how the momentum-space Riemann curvature could be extracted from the wavepacket dynamics, on setting controllable electric field strengths.

We begin with the geodesic equation in momentum space and for the purposes of this section, set $e = \hbar = 1$. The simplified final result of Appendix~\ref{app:A} then reads,
%
\begin{align}
    \label{eq:geodesic3}
    &(\dot{\vec{r}}_c)_\gamma = \partial_{k_\gamma} \tilde{\varepsilon}_{n \vec{k}} \\&+ \Big( \partial_{k_\gamma} \mathcal{G}_{\alpha \beta}^n(E^2) - \frac{1}{2} [\partial_{k_\alpha} \mathcal{G}_{\beta \gamma}^n(E^2) + \partial_{k_\beta} \mathcal{G}_{\alpha \gamma}^n(E^2)] \Big) \dot{k}_\alpha \dot{k}_\beta,\nonumber
\end{align}
\\
which using the definition of a Christoffel symbol from the main text reduces to,
\beq{}
    \label{eq:geodesic4}
    (\dot{\vec{r}}_c)_\gamma = \partial_{k_\gamma} \tilde{\varepsilon}_{n \vec{k}} + \Big( \frac{1}{2} \partial_{k_\gamma} \mathcal{G}_{\alpha \beta}^n(E^2) -  \Gamma_{\gamma \alpha \beta}^n(E^2) \Big) \dot{k}_\alpha \dot{k}_\beta.
\eeq{}

We further utilize the definition of the weighted momentum-space Riemann tensor introduced in the main text,
\beq{}
    \label{eq:riemann}
    \mathcal{R}^n_{\alpha \beta \gamma \delta} = \partial_{k_\delta} \Gamma^n_{\alpha \beta \gamma } - \partial_{k_\gamma} \Gamma^n_{\alpha \beta \delta} + \Gamma^n_{ \mu \alpha \delta} \Gamma^n_{\beta \mu \gamma} - \Gamma^n_{\nu \beta \delta} \Gamma^n_{\alpha \nu \gamma}
\eeq{}
where $\mu$ and $\nu$ are dummy indices. On differentiating Eq.~\eqref{eq:geodesic4} with respect to $k_\delta$, we obtain
\begin{widetext}
\begin{align}
    \partial_{k_\delta} (\vec{\dot{r}}_c)_\gamma &= \partial_{k_\delta} \partial_{k_\gamma} \tilde{\varepsilon}_{n \vec{k}} + \partial_{k_\delta} \Big[\frac{1}{2} \partial_{k_\gamma} \mathcal{G}_{\alpha \beta}^n -  \Gamma_{\gamma \alpha \beta}^n\Big]\dot{k}_\beta \dot{k}_\alpha  + \Big[ \frac{1}{2} \partial_{k_\gamma} \mathcal{G}_{\alpha \beta}^n -  \Gamma_{\gamma \alpha \beta}^n\Big]\partial_{k_\delta} (\dot{k}_\beta \dot{k}_\alpha) \nonumber \\
    &= \partial_{k_\delta} \partial_{k_\gamma} \tilde{\varepsilon}_{n \vec{k}} + \partial_{k_\delta} \Big[ \frac{1}{2} \partial_{k_\gamma} \mathcal{G}_{\alpha \beta}^n -  \Gamma_{\gamma \alpha \beta}^n\Big] \dot{k}_\beta \dot{k}_\alpha + \Big[ \frac{1}{2} \partial_{k_\gamma} \mathcal{G}_{\alpha \beta}^n -  \Gamma_{\gamma \alpha \beta}^n \Big]( \dot{k}_\alpha(\partial_{k_\delta} \partial_t k_\beta) +  \dot{k}_\beta(\partial_{k_\delta} \partial_t k_\gamma)) \nonumber \\
    &= \partial_{k_\delta} \partial_{k_\gamma} \tilde{\varepsilon}_{n \vec{k}} + \partial_{k_\delta} \Big[ \frac{1}{2} \partial_{k_\gamma} \mathcal{G}_{\alpha \beta}^n -  \Gamma_{\gamma \alpha \beta}^n\Big] \dot{k}_\beta \dot{k}_\alpha + \Big[ \frac{1}{2} \partial_{k_\gamma} \mathcal{G}_{\alpha \beta}^n -  \Gamma_{\gamma \alpha \beta}^n\Big](\dot{k}_\alpha \partial_t \delta_{\delta \beta} + \dot{k}_\beta \partial_t \delta_{\delta \gamma}) \nonumber \\
    &= \partial_{k_\delta} \partial_{k_\gamma} \tilde{\varepsilon}_{n \vec{k}} + \frac{1}{2} \partial_{k_\delta} \partial_{k_\gamma} \mathcal{G}_{\alpha \beta}^n \dot{k}_\beta \dot{k}_\alpha -  \partial_{k_\delta} \Gamma_{\gamma \alpha \beta}^n \dot{k}_\beta \dot{k}_\alpha,
\end{align}

where we have used the fact that the partial derivatives commute. Antisymmetrizing with respect to $\gamma$ and $\delta$, and using again the  commutativity of derivatives, the first term on the right-hand-side cancels out.
Using Eq.~\eqref{eq:riemann} and rearranging yields
\beq{}
    \frac{1}{\dot{k}_\beta \dot{k}_\alpha} \Big( \partial_{k_\delta} (\dot{\vec{r}_c})_\gamma - \partial_{k_\gamma} (\dot{\vec{r}_c})_\delta \Big) = \partial_{k_\delta} \Gamma_{\gamma \alpha \beta}^n - \partial_{k_\gamma} \Gamma_{\delta \alpha \beta}^n  = \mathcal{R}^n_{\alpha \beta \gamma \delta} - \Gamma_{\beta \mu \delta} \Gamma_{\mu \gamma \alpha} + \Gamma_{\beta \nu \alpha} \Gamma_{\nu \gamma \delta}.
\eeq
On substituting the values of the appropriate Christoffel symbols from Eq.~\eqref{eq:geodesic4}, this reads

\beq{}
\begin{split}
    \frac{1}{\dot{k}_\beta \dot{k}_\alpha} \Big(\partial_{k_\delta}  (\dot{\vec{r}}_c)_\gamma - \partial_{k_\gamma}  (\dot{\vec{r}}_c)_\delta\Big) = \mathcal{R}^n_{\alpha \beta \gamma \delta} &- \frac{[(\dot{\vec{r}}_c)_\delta - \partial_{k_\delta} \tilde{\varepsilon}_{n \vec{k}}- \frac{1}{2}  \partial_{k_\delta} \mathcal{G}_{\beta \mu}^n]}{ \dot{k}_\beta \dot{k}_\mu}\frac{[(\dot{\vec{r}}_c)_\gamma - \partial_{k_\gamma} \tilde{\varepsilon}_{n \vec{k}}- \frac{1}{2} \partial_{k_\gamma} \mathcal{G}_{\alpha \mu}^n]}{\dot{k}_\mu \dot{k}_\alpha}\\
    &+ \frac{[(\dot{\vec{r}}_c)_\gamma - \partial_{k_\gamma} \tilde{\varepsilon}_{n \vec{k}} - \frac{1}{2}  \partial_{k_\gamma} \mathcal{G}_{\beta \nu}^n]}{ \dot{k}_\beta \dot{k}_\nu}\frac{[(\dot{\vec{r}}_c)_\delta - \partial_{k_\delta} \tilde{\varepsilon}_{n \vec{k}}-\frac{1}{2} \partial_{k_\delta} \mathcal{G}_{\nu \alpha}^n ]}{\dot{k}_\nu \dot{k}_\alpha}.
\end{split}
\eeq{}
\end{widetext}
The last two terms cancel (since both dummy indices $\mu$ and $\nu$ sum over to $k^2$). This leaves us with the compact result for the deviation equation
\beq{}
     \partial_{k_\delta} (\dot{\vec{r}}_c)_\gamma - \partial_{k_\gamma} (\dot{\vec{r}}_c)_\delta = \mathcal{R}^n_{\alpha \beta \gamma \delta} \dot{k}_\alpha \dot{k}_\beta.
\eeq{}
The physical interpretation of the geodesic equation derived here is the following. The Riemannian curvature $\mathcal{R}^n_{\alpha \beta \gamma \delta}$ determines the differences in the wavepacket evolutions governed by the center-of-mass velocities $\dot{\vec{r}}_c$, given their initial center-of-mass momenta $\vec{k}_c$. In particular, on retrieving $e$, $\hbar$, and inserting $\dot{\vec{k}}_c = -e\frac{\vec{E}}{\hbar}$, we have
\beq{}
     \partial_{k_\delta} (\dot{\vec{r}}_c)_\gamma - \partial_{k_\gamma} (\dot{\vec{r}}_c)_\delta = \frac{e^2}{\hbar^2} \mathcal{R}^n_{\alpha \beta \gamma \delta} E_\alpha  E_\beta,
\eeq{}
which demonstrates that the Riemann curvature $\mathcal{R}^n_{\alpha \beta \gamma \delta}$ can be reconstructed from wavepacket deviations, on controlling the electric field components $\vec{E} = (E_x, E_y)$.


\section{Numerical details for obtaining wavepacket trajectories}\label{app:C}

Here we provide further numerical details for obtaining the wavepacket trajectories demonstrated in Fig.~\ref{Fig:Wavepackets} of the main text.
The systems were initialized with a three-band gapless kagome Hamiltonian as described in Sec.~\ref{sec:III}~\cite{Jiang2021}, with parameters $\varepsilon_A = \varepsilon_B = \varepsilon_C = 0$, $t=1$, and $t' = -0.2$ on a hexagonal lattice with primitive reciprocal vectors $k_1, k_2 = \pm \frac{3}{2} k_x + \frac{\sqrt{3}}{2} k_y$. The eigenvectors of this Hamiltonian were used to calculate the multiband quantum metrics $g^{nm}_{\alpha \beta}$ via the non-Abelian Berry connections for all ordered band pairs. The weighted metrics $\mathcal{G}^{nm}_{\alpha \beta}$ for the nonuniform electric field and nonlinear field coupling were then found, which in turn yielded the field-induced Berry curvature corrections $\Tilde{\Omega}_n(\vec{E})$. The wavepacket evolution was subsequently obtained via simultaneous numerical solution of the semiclassical equation of motion [Eq.~\eqref{eq:derivative_semiclassical} for nonuniform fields and Eq.~\eqref{eq:kdot} for nonlinear coupling] for position and momentum as a function of time. All wavepackets were initialized at the origin in real space, but at several different momenta in the BZ, in order to probe the effects of proximity to the nodes at the $\Gamma$ and $K$ points. The numerical solver was implemented in Wolfram Mathematica 14.1 with the \textit{Residual} method where a DAE solver was used with maximum time step size $\Delta t = 0.1$. The resultant numerical solutions demonstrating the anomalous Euler wavepacket propagation induced by the geodesic (Appendix~\ref{app:A}) and geodesic deviation equations (Appendix~\ref{app:B}) are shown in Fig.~\ref{Fig:Wavepackets} of the main text.

\section{Numerical momentum-space horizons}\label{app:D}

\begin{figure}
    \centering
    \includegraphics[width=\columnwidth]{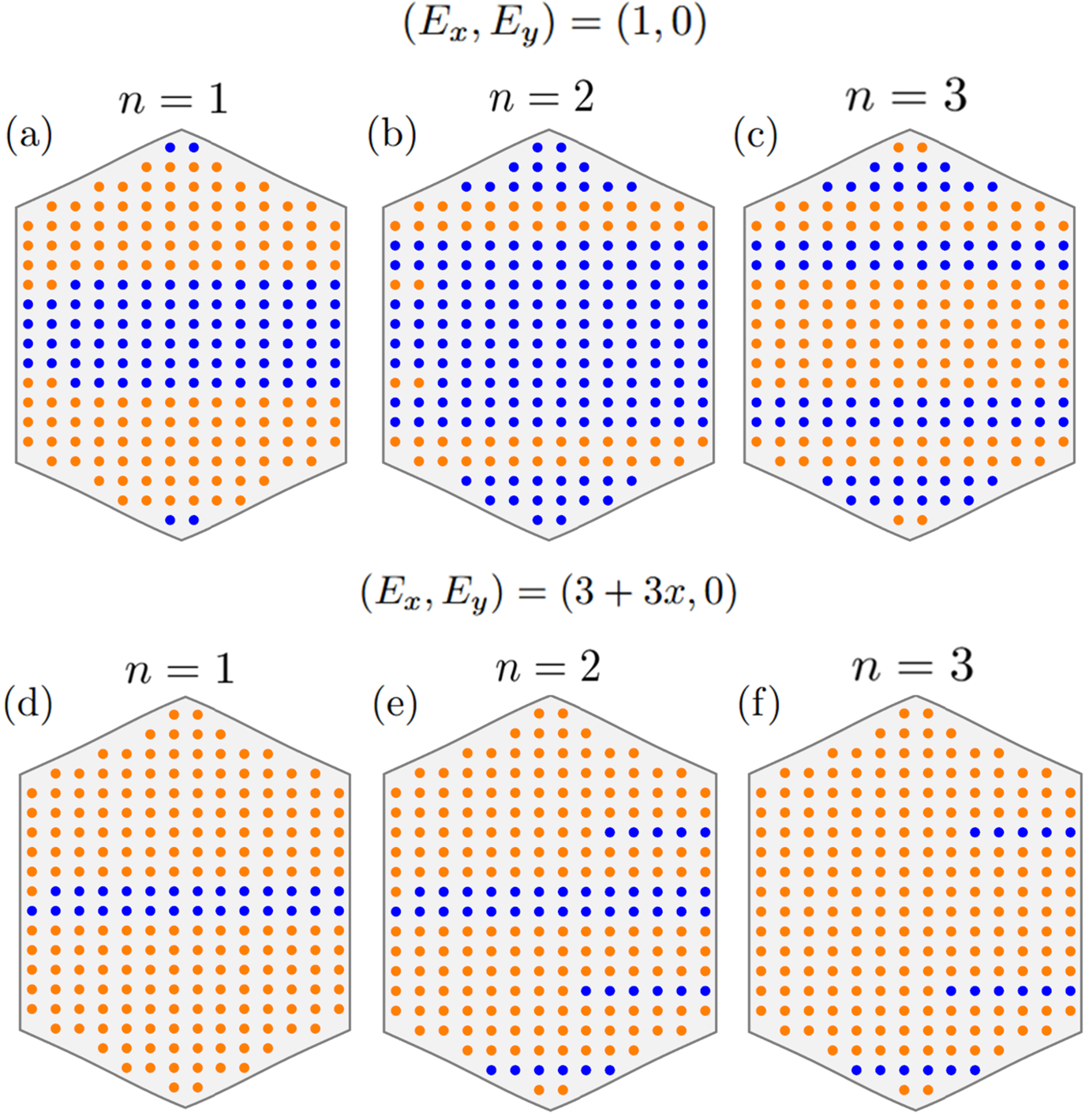}
    \caption{Momentum-space horizons induced by the singular quantum metric, numerically retrieved from identifying the momenta of wavepackets falling into the band nodes. Individual points represent the initial wavevector $\vec{k}_0$ of a wavepacket and blue (orange) color represents divergent (convergent) evolution. Boundaries between regions show the horizons. \textbf{(a)}--\textbf{(c)} Horizon plots on nonlinear quadratic coupling to a uniform electric field $(E_x, E_y) = (1,0)$ for the three energy bands. \textbf{(d)}-\textbf{(f)} Horizon plots on coupling to a constant electric field gradient $(E_x, E_y) = (3+3x,0)$ for the three energy bands.}
    \label{fig:horizons}
\end{figure}

We describe here the evaluation of the momentum space horizons referred to in the main text. Correspondingly, we use the field-induced Berry curvature corrections $\Tilde{\Omega}_n(\vec{E})$ obtained in the aforementioned wavepacket evolution simulation (see Appendices~\ref{app:A} and~\ref{app:C}). A regular mesh of initial momenta $\vec{k}_0$ in the BZ [$(-\pi,\pi) \times (-\pi,\pi)$] with spacing $\Delta k = 0.3$ was subjected to the same DAE solver. Evolution into the quantum-metric singularity was tabulated for each point, in both nonuniform fields and nonlinear coupling, depending on whether the wavepacket trajectories evolved to divergent distances from the origin due to the singular geometric term, within the time interval probed. We present the tabulated result in Fig.~\ref{fig:horizons}. The results were verified for different spacings and electric field directions, where we obtained analogous horizons satisfying the symmetries of the kagome lattice that were explicitly broken with the external electric fields that drive the wavepacket evolution. Notably, the combined momentum-space horizons within bands $n=1$ and $n=3$ overlap with the horizons obtained in band $n = 2$, consistently with the band dispersion and multiband quantum geometry of two nodal subspaces realized between three bands.

\section{Riemannian geometry of an Euler node}\label{app:E}
\begin{figure}
    \centering
    \includegraphics[width=\linewidth]{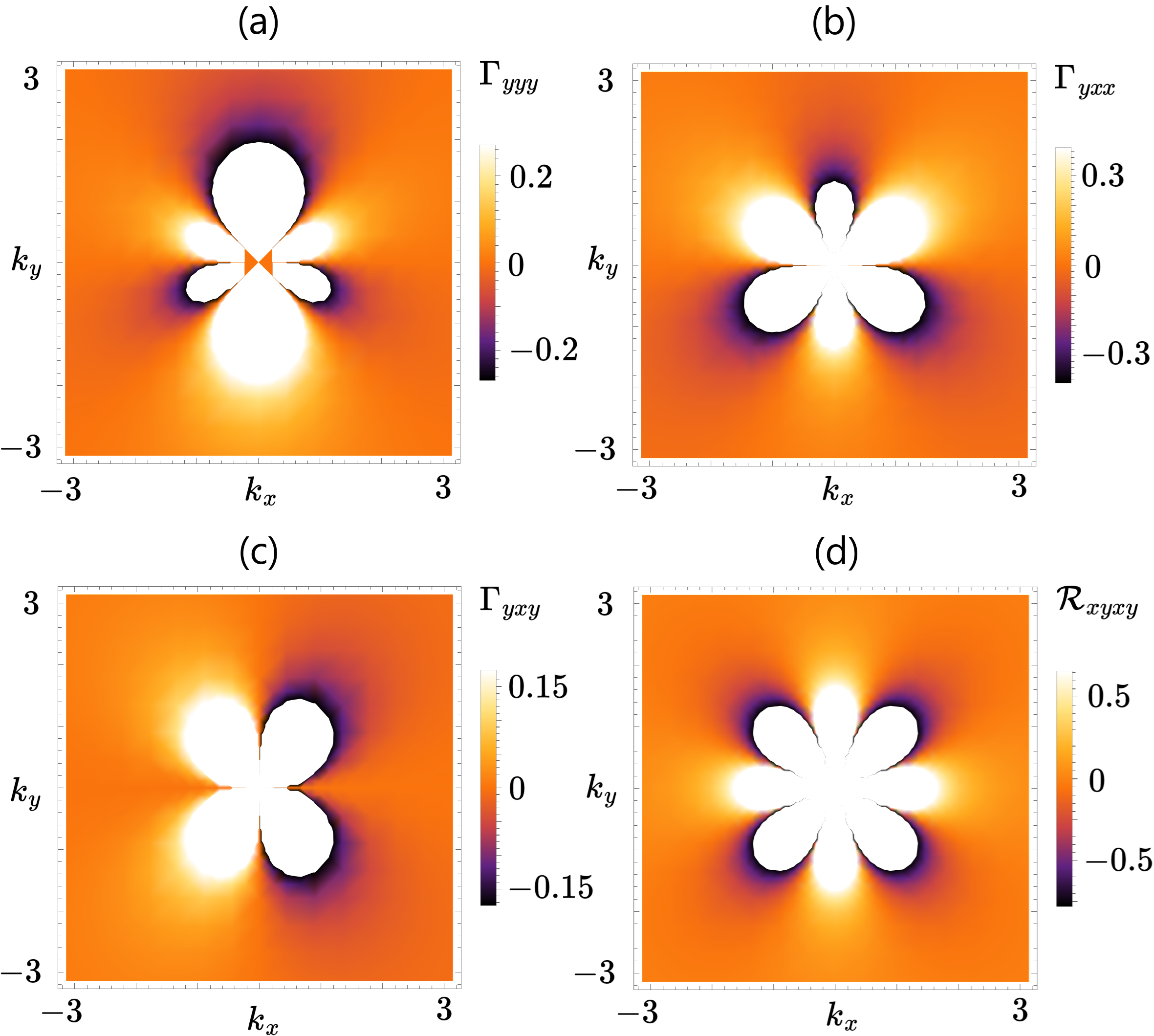}
    \caption{Riemannian geometry of the continuum models with topological Euler class. We demonstrate the momentum space dependence of Christoffel symbols that enter the wavepacket geodesic equation (see Appendix~\ref{app:A}) and the anomalous second-order Hall responses. \textbf{(a)} $\Gamma^{12}_{yyy}$, \textbf{(b)} $\Gamma^{12}_{yxx}$, \textbf{(c)} $\Gamma^{12}_{yxy}$. \textbf{(d)} Momentum-space Riemann curvature $\mathcal{R}^{12}_{xyxy}$. $\mathcal{R}^{12}_{xyxy}$ reflects the nonvanishing anomalous third-order conductivity due to second derivatives of quantum-metric, which were here induced by the Euler node with topological charge $\chi = 1$.}
    \label{Fig:Riemann}
\end{figure}

In the following, we fully characterize the Riemannian geometry of a non-Abelian band singularity hosting Euler class. The continuum Hamiltonian of an Euler node reads~\cite{BJY_nielsen, Morris_2024, jankowski2023optical}
\beq{eq:EulerNode}
H_\chi(\kv)=\alpha(k_+^{2|\chi|}\sigma_-+k^{2|\chi|}_-\sigma_+),
\eeq
with constant $\alpha$ fixing the band dispersion, $k_{\pm}=k_x \pm i k_y$, and ${\sigma_{\pm}=(\sigma_x\pm i\sigma_z)/2}$, which manifestly satisfies the $\mathcal{C}_2 \mathcal{T}$ symmetry, allowing to write the Hamiltonian in a real gauge $[H(\kv) = H^*(\kv)]$, with real eigenvectors. Real eigenvectors define a real vector bundle over the momentum space, which given the reality of the Bloch bundle, can be topologically characterized with the Euler characteristic class $\chi \in \mathbb{Z}$. As introduced in the main text, the Euler class $\chi$ is associated with a singular non-Abelian connection $\vec{A}_{12} = i \vec{a} = i\bra{u_{1\kv}}\ket{\nabla_\kv u_{2\kv}}$, which, as a direct consequence, induces a singular quantum metric $g^{12}_{\alpha \beta}$. From the eigenvectors of the Hamiltonian $H(\kv)$, that under the present symmetry can be chosen real, we directly evaluate the non-Abelian Berry connection and the quantum metric components~\cite{jankowski2023optical},
\beq{}
   g^{12}_{\alpha \beta} = A^{\alpha}_{12} A^{\beta}_{21} = \frac{\chi^2 k_\alpha k_\beta}{k^4} (2\delta_{\alpha \beta}-1),
\eeq
with $k^2 \equiv k^2_x + k^2_y$. On differentiating in momentum-space parameters $k_x, k_y$, we further obtain the momentum-space Christoffel symbols~\cite{Ahn2020, Ahn2021, Hetenyi2023, jankowski2023optical} relevant to the nonlinear transport
\beq{}
    \Gamma^{12}_{xxx} = \frac{1}{2} \partial_{k_x} g^{12}_{xx} = \frac{-2\chi^2 k_x k^2_y}{k^6},
\eeq
\beq{}
    \Gamma^{12}_{yyy} = \frac{1}{2} \partial_{k_y} g^{12}_{yy} = \frac{-2\chi^2 k^2_x k_y}{k^6},
\eeq
\beq{}
    \Gamma^{12}_{yxx} = \partial_{k_x} g^{12}_{xy} - \frac{1}{2} \partial_{k_y} g^{12}_{xx} = \frac{\chi^2}{k^6} (3k^3_y - k^2_x k_y - k_x k^2_y + k^3_x),
\eeq
\beq{}
    \Gamma^{12}_{xyy} =  \partial_{k_y} g^{12}_{yx} - \frac{1}{2} \partial_{k_x} g^{12}_{yy} = \frac{\chi^2}{k^6} (3k^3_x - k^2_y k_x - k_y k^2_x + k^3_y),
\eeq
\beq{}
    \Gamma^{12}_{xxy} = \Gamma^{12}_{xyx} = \frac{1}{2} \partial_{k_y} g^{12}_{xx} = \frac{\chi^2 (k^2_x - k^2_y)k_y}{k^6},
\eeq
\beq{}
    \Gamma^{12}_{yyx} = \Gamma^{12}_{yxy} = \frac{1}{2} \partial_{k_x} g^{12}_{yy} = \frac{\chi^2 (k^2_y - k^2_x)k_x}{k^6}.
\eeq
Finally, we compute the Riemann tensor within the continuum model, obtaining a single nontrivial component 
%
\begin{multline}
      \mathcal{R}^{12}_{xyxy} = \partial_{k_x} \Gamma^{12}_{xyy} - \partial_{k_y} \Gamma^{12}_{xyx} + \Big( \Gamma^{12}_{xxx} \Gamma^{12}_{xyy} - \Gamma^{12}_{xxy} \Gamma^{12}_{xxy} \Big) \\ + \Big( \Gamma^{12}_{xyx} \Gamma^{12}_{yyy} - \Gamma^{12}_{xyy} \Gamma^{12}_{yxy} \Big),
\end{multline}
%
where we use a general definition
\beq{}
\mathcal{R}^{12}_{\alpha \beta \gamma \delta} = \partial_{k_\gamma} \Gamma^{12}_{\alpha \beta \delta} - \partial_{k_\delta} \Gamma^{12}_{\alpha \beta \gamma} + \sum_\mu \Big( \Gamma^{12}_{\alpha \mu \gamma} \Gamma^{12}_{\mu \beta \delta} - \Gamma^{12}_{\alpha \mu \beta} \Gamma^{12}_{\mu \gamma \delta} \Big).
\eeq
Correspondingly, we obtain
\beq{}
    \mathcal{R}^{12}_{xyxy} = \frac{\chi^2 (5 k^4_x + 5 k^4_y - 38 k^2_x k^2_y)}{k^8},
\eeq
which further allows to compute the Kretschmann (scalar) invariant~\cite{K1915},
\beq{}
    K = 6 (\mathcal{R}^{12}_{xyxy})^2 = \frac{6 \chi^4 (5 k^4_x + 5 k^4_y - 38 k^2_x k^2_y)^2}{k^{16}}.
\eeq
Manifestly, the momentum space Kretschmann invariant diverges as $K \sim |k|^{-8}$ at $\vec{k} = 0$, which corresponds to the quantum metric singularity at the momentum-space position of the Euler node. We demonstrate the singular nature of the momentum-space Christoffel symbols and momentum-space Riemann curvature in Fig.~\ref{Fig:Riemann}.

\section{Reconstructing the Euler invariant with electric fields}\label{app:F}

In the following, we outline how the Euler invariant can be reconstructed from the measurements of electric current responses to nonuniform electric fields and from the second-order currents in response to electric fields. We consider two reconstruction schemes from: (1) anomalous Euler wavepacket dynamics in nonlinear response to the electric fields and electric field gradients; (2) bulk anomalous (Hall) currents presented in the main text.

(1) In the case of wavepackets, deducing the center-of-mass trajectories amounts to an integration of the geodesic equation derived in Appendix~\ref{app:A},
%
\begin{align}\label{eq:extract}
    &(\vec{r}_c)_\gamma (t) = \int^t_0 \dd t'~ (\dot{\vec{r}}_c)_\gamma (t') = \frac{1}{\hbar} \int^t_0 \dd t'~  \partial_{k_\gamma} \varepsilon_{n \vec{k}} \\ &+ \frac{e^2}{\hbar^2}\int^t_0 \dd t'~ \Big( \frac{1}{2} \partial_{k_\gamma} \mathcal{G}_{\alpha \beta}^n(E^2) -  \Gamma_{\gamma \alpha \beta}^n(E^2) \Big) E_\alpha (t') E_\beta (t'). \nonumber 
\end{align}
%
Therefore, adiabatic changes to the electric fields $E_\alpha (t')$ could be used to extract the values of the linear combinations of the derivatives $\partial_{k_\gamma} \mathcal{G}_{\alpha \beta}^n(E^2)$ present in Eq.~\eqref{eq:extract}. Similarly, changing the electric field gradients adiabatically on top of the uniform electric fields present, allows to deduce the deflections of initial trajectories,
\beq{}
     (\Delta \vec{r}_c)_\gamma (t) = \int^t_0 \dd t'~\Big( \frac{e}{2\hbar} \partial_{k_\gamma} \mathcal{G}_{\alpha \beta}^n(\nabla) \partial_\alpha E_\beta (t') \Big).
\eeq
In Fig.~\ref{Fig:Wavepackets}, we precisely demonstrate the alteration of the wavepacket trajectories subject to the addition of electric field gradients. The gradient-induced deflections allow to extract the derivatives $\partial_{k_\gamma} \mathcal{G}_{\alpha \beta}^n(\nabla)$, hence the derivatives of the quantum metric are extracted, on an identification $\mathcal{G}_{\alpha \beta}^n(\nabla) = g^n_{\alpha \beta}$. To extract the metric itself, one can asymptotically continue, i.e., interpolate $\partial_{k_\gamma} g^n_{\alpha \beta}$ to ${k \rightarrow \infty}$, where ${g^n_{\alpha \beta} = 0}$, consistently with the continuum models (see Appendix~\ref{app:E}). This fixes the integration constant for integrating the derivatives to obtain the metric: $g^n_{\alpha \beta}(\kv) = \int^\kv_\infty \dd \kv' \cdot \nabla_{\kv'} g^n_{\alpha \beta}$, where we perform a set of line integrals. An analogous procedure obtains ${\mathcal{G}_{\alpha \beta}^n(E^2) = G^n_{\alpha \beta}}$, once the individual derivatives, $\partial_{k_\gamma} \mathcal{G}_{\alpha \beta}^n(E^2)$, are disentangled within the previously retrieved linear combinations. To obtain individual derivatives $\partial_{k_\gamma} \mathcal{G}_{\alpha \beta}^n(E^2)$, one can assume that the derivatives $\partial_{k_\gamma}
G^n_{xy}$ are much smaller than $\partial_{k_\gamma}
G^n_{\alpha \alpha}$ (${\partial_{k_\gamma}
G^n_{xy} \ll \partial_{k_\gamma}
G^n_{\alpha \alpha}}$), consistently with the findings of negligible $\partial_{k_\gamma}
G^n_{xy}$ contributions to the anomalous wavepacket dynamics demonstrated in Fig.~\ref{Fig:Wavepackets}, and consistently with the negligibly small values of $\sigma_{y,xy}(\nabla)$ in Fig.~\ref{Fig:Bulk}. The neglected values of $\partial_{k_\gamma}
G^n_{xy}$, can nevertheless be retrieved exactly, on achieving a self-consistency with the numerically obtained non-Abelian Berry connection elements, which we retrieve further after obtaining the weighted quantum metrics within the first approximation. If the derivatives $\partial_{k_\gamma}
G^n_{xy}$, $\partial_{k_\gamma}
G^n_{\alpha \alpha}$ were comparable (${\partial_{k_\gamma}
G^n_{xy} \sim \partial_{k_\gamma}
G^n_{\alpha \alpha}}$), and no self-consistency within the approximation could be achieved, the quadratic coupling results can be augmented with the derivative coupling results, where no linear combinations are present from the beginning. In that case, the ratios $\mathcal{G}^n_{xy}(\nabla)/\mathcal{G}^n_{\alpha\alpha}(\nabla)$ allow to estimate the local ratios ${\partial_{k_\gamma}
G^n_{xy} / \partial_{k_\gamma}
G^n_{\alpha \alpha}}$ to simplify the linear combinations over the $k$-space, which serve as an alternative starting point for achieving the self-consistency with the values of multiband metrics and connections obtained afterwards.

Furthermore, on knowing the band energies, as e.g., experimentally accessible from ARPES measurements, we evaluate the weights $c_{nm} = \hbar/(\varepsilon_{n\kv}-\varepsilon_{m\kv})$. Hence, for every $\kv$-point, we arrive at a set of equations $G_{\alpha \beta}^n = \sum_{m \neq n} c_{nm} g^{nm}_{\alpha \beta}$, which requires to remain consistent with $g_{\alpha \beta}^n = \sum_{m \neq n} g^{nm}_{\alpha \beta}$. Although this protocol allows to reconstruct $g^{n}_{\alpha \beta}$, it does not provide the full multiband resolution in $g^{nm}_{\alpha \beta}$. However, given the singular nature of the Euler nodes, one could expect that close to the band node momentum: $g^{n}_{\alpha \beta} \approx g^{n (n+1)}_{\alpha \beta} = A^\alpha_{n.n+1} A^\beta_{n+1,n}$, which on combining all metric coefficients $a,b = x,y$, allows to deduce $A^\alpha_{n.n+1}$ up to a sign, similarly to probing the optical transition matrix elements with light~\cite{jankowski2023optical}. As shown in Ref.~\cite{jankowski2023optical}, the Euler invariant $\chi$ can be deduced on maximally-smooth $\mathbb{Z}_2$ (sign) gauge fixing, which allows to reconstruct the vorticity in non-Abelian Berry connection, which defines the Euler class, as outlined in the main text.

(2) In the case of \textit{bulk currents}, the Euler class could be inferred from the scaling of the current divergence on doping the Euler nodes. Changing $\mu$ allows to infer the contributions to anomalous conductivities $\sigma_{\gamma, \alpha \beta}(\nabla)$, $\sigma_{\gamma, \alpha \beta}(E^2)$, $\sigma_{\delta, \alpha \beta \gamma}(E^3)$, from averages of $\langle \partial_{k_\gamma} G^n_{\alpha \beta} \rangle_\mu$, $\langle \partial_{k_\gamma} g^n_{\alpha \beta} \rangle_\mu$ at given energies $E = \mu$. Similarly to the wavepacket case, this allows to estimate a radial dependence of averaged $G^n_{\alpha \beta}$, $g^n_{\alpha \beta}$ from the node, which in turn allows to estimate the magnitude and scaling of $A^\alpha_{n,n+1}$. A radial estimate of $A^\alpha_{n,n+1}$ allows to predict the Euler class, given the non-singular nodes (with $\chi = 0$) should realize vanishing averages $A^\alpha_{n.n+1}$, as the vorticity in non-Abelian Berry connection is not present in such case.

\section{Derivation of the higher-order bulk Hall currents}\label{app:G}

\begin{figure}
    \centering
    \includegraphics[width=\linewidth]{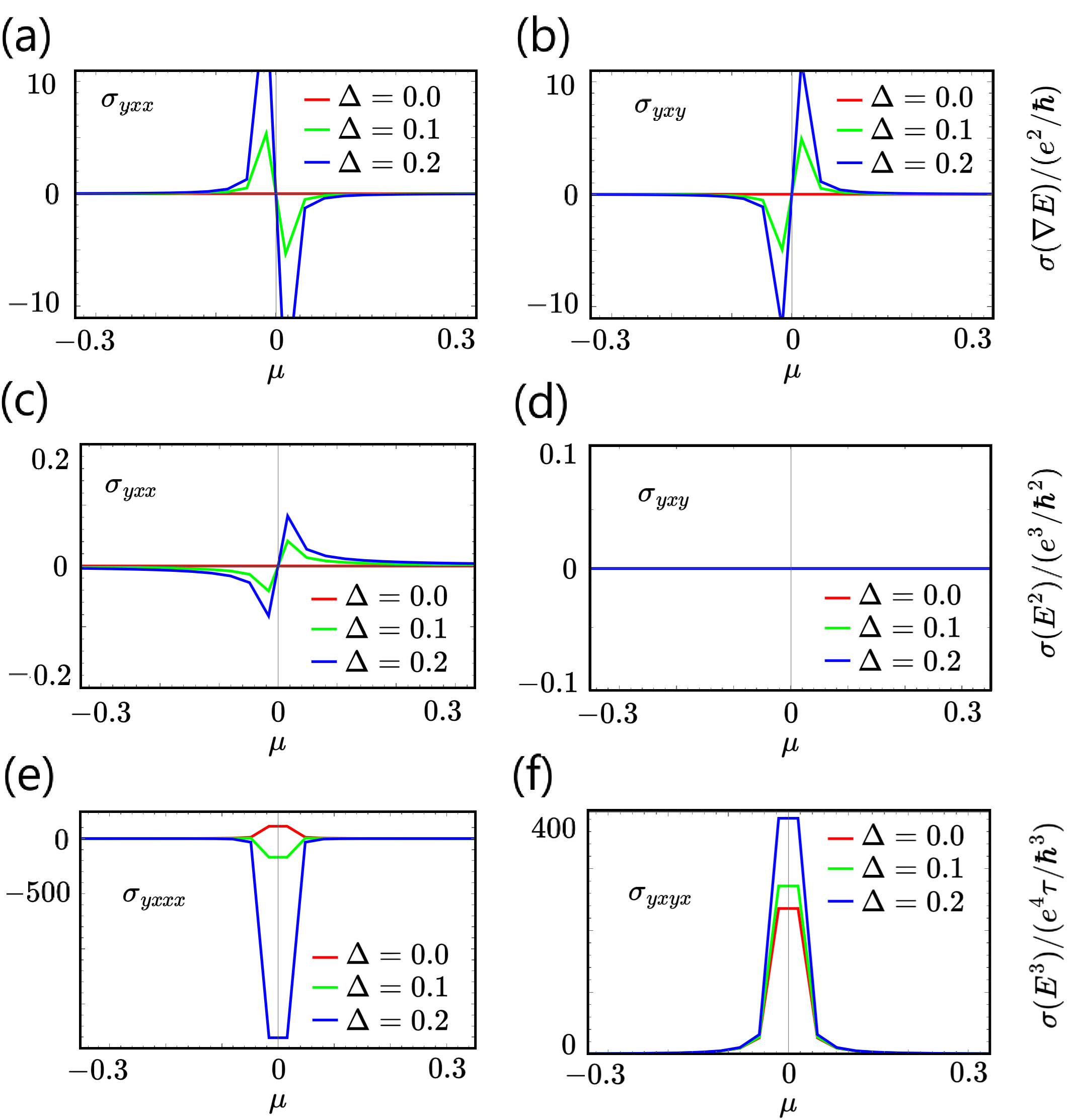}
    \caption{Anomalous currents in continuum models with topological Euler class, as a function of chemical potential $\mu$ and inversion-symmetry breaking perturbations of strength~$\Delta$. For analogous components of conductivity tensors within the lattice-regularized models, see Fig.~\ref{Fig:Bulk} in the main text. \mbox{\textbf{(a)}--\textbf{(b)}} Conductivities yielding geometric currents in responses to electric field gradients. \textbf{(c)}--\textbf{(d)}  Quantum-metric dipole-induced second-order Hall conductivities in responses to electric fields. \textbf{(e)}--\textbf{(f)} Quantum-metric quadrupole-induced third-order Hall conductivities in responses to electric fields. The third-order geometric Hall response is extrinsic and directly proportional to the scattering time $\tau$.}
    \label{Fig:contCond}
\end{figure}

For completeness, we derive the quantum-geometric contributions to second-order and third-order electric conductances in response to electric fields. A detailed study of third-order contributions was provided by Ref.~\cite{Mandal2024}. The general approach relies on deducing the perturbative response of the free fermion eigenstates and eigenenergies, as concerning the semiclassical wavepacket dynamics (see Appendix~\ref{app:A}). In addition, one needs to renormalize the occupation factors of the perturbed eigenstates within the kinetic Boltzmann transport formalism,
\beq{}
    \vec{j} = -e \int_\kv \sum_n f_{n\kv} \vec{v}_{n \kv} = -e \int_{\kv_c} \sum_n f_{n}(\vec{r}_c, \vec{k}_c, t) \dot{\vec{r}}_c.
\eeq
The velocity operator matrix elements $\vec{v}_{n \kv}$ can be deduced within the approach of Appendix~\ref{app:A}, where we explicitly found the wavepacket velocities $\dot{\vec{r}}_c$ for any band $n$. The distribution function $f_{n}(\vec{r}_c, \vec{k}_c, t)$ can be perturbatively expanded as $f_{n} = f^{(0)}_n + f^{(1)}_n + f^{(2)}_n + \ldots$, where $f^{(0)}_n = f_{n\kv} (\mu)$ is the equilibrium Fermi-Dirac distribution, as specified in the main text. The Boltzmann transport equation reads
\beq{}
    \partial_t f_n + \dot{\vec{k}}_c \cdot \nabla_\kv f_n + \dot{\vec{r}}_c \cdot \nabla_\rv f_n = -\frac{f_n-f^{(0)}_n}{\tau},
\eeq
where the right-hand side corresponds to the scattering with scattering time $\tau$. Inserting, $\dot{\vec{k}}_c = -\frac{e \vec{E}(\vec{r}_c, t) }{\hbar}$, with $\vec{E}(\vec{r}_c, t) = \text{Re}~(\vec{E} e^{i \om t})$ yields a hierarchy in frequency domain,
\beq{}
    i \om f^{(j+1)}_n + \frac{e \vec{E}}{\hbar} \nabla_\kv f^{(j)}_n = -\frac{f^{(j+1)}_n}{\tau}
\eeq
which amounts to distribution corrections in the powers of electric field strength
\beq{}
    f^{(j)}_n = \Big(-\frac{e}{\hbar (i \om + 1/\tau)}\Big)^j (\vec{E} \cdot \nabla_\kv)^j f^{(0)}_n.
\eeq
For dc responses central to this work, we send $\om \rightarrow 0$.

Equipped with the corrections in the distribution function, we now evaluate quantum-geometric contributions to second and third-order currents,
At second order, the quantum geometric term~\cite{Gao2014} on combining wavepacket velocities and perturbed distribution yields:
\beq{}
    \vec{j}^{(2)} =  -e \int_{\kv} \sum_n f^{(0)}_{n} \Big( - \frac{e}{\hbar} \vec{E} \times  [\nabla_\kv \times (\underline{\underline{\mathcal{G}^n}}(E^2) \vec{E}) ] \Big),
\eeq
which simplifies to~\cite{Kaplan2024},
%
\begin{multline}
    j^{(2)}_\gamma = -\frac{e^3}{\hbar^2} E_\alpha E_\beta  \int_{\kv} \sum_n f_{n \kv}(\mu) \Big( 2 \partial_{k_\gamma} \mathcal{G}_{\alpha \beta}^n(E^2) \\ - \frac{1}{2} [ \partial_{k_\alpha} \mathcal{G}_{\beta \gamma}^n(E^2) + \partial_{k_\beta} \mathcal{G}_{\alpha \gamma}^n(E^2) ] \Big).   
\end{multline}
%
Manifestly, $\vec{j}^{(2)} \cdot \vec{E} = 0$, showing the dissipationless (Hall) nature of the response. We calculate the corresponding second-order quantum-geometric Hall conductivity,
%
\begin{multline}
   \sigma^{\text{H}}_{\gamma, \alpha \beta}(E^2) = -\frac{e^3}{\hbar^2} \int_{\kv} \sum_n f_{n \kv}(\mu) \Big( \partial_{k_\gamma} \mathcal{G}_{\alpha \beta}^n(E^2) \\ - \frac{1}{2} [\partial_{k_\alpha} \mathcal{G}_{\beta \gamma}^n(E^2) + \partial_{k_\beta} \mathcal{G}_{\alpha \gamma}^n(E^2) ] \Big)
\end{multline}
%
for Euler Hamiltonians in the main text. Importantly, while this nonlinear Hall conductivity admits $\mathcal{C}_2\mathcal{T}$ symmetry, it requires $\mathcal{C}_2$ and $\mathcal{T}$ symmetries individually to vanish~\cite{Gao2014}.

The other nonlinear Hall contribution at second order in electric fields~\cite{Sodemann2015},
%
\begin{multline}    
    \vec{j}^{(2)} =  -e \int_{\kv} \sum_n f^{(1)}_{n} \Big( - \frac{e}{\hbar} [\vec{E} \times  \vec{\Om}_n ] \Big) \\= -\frac{e^3 \tau}{\hbar} E_\alpha E_\beta \int_{\kv} \sum_n f^{(0)}_{n} \Big( \partial_{k_\alpha} \Om^n_{\beta \gamma} + \partial_{k_\beta} \Om^n_{\alpha \gamma} \Big),
\end{multline}
%
associated with the Berry curvature dipole (on integrating by parts), vanishes in the Euler Hamiltonians, as $\vec{\Omega}_n$ itself vanishes identically. On inserting $f^{(2)}_{n}$, one could analogously derive the third-order Hall Berry curvature dipole response~\cite{Mandal2024}, which also vanishes in the $\mathcal{C}_2\mathcal{T}$-symmetric systems with nontrivial Euler class.

The only nonvanishing third-order Hall contribution compatible with $\mathcal{C}_2\mathcal{T}$, $\mathcal{C}_2$, and $\mathcal{T}$ symmetries in centrosymmetric Euler Hamiltonians arises from the quantum metric quadrupole response, which unlike the intrinsic (i.e., $\tau^0$ in the scattering time) quantum metric dipole Hall effect, is extrinsic, and arises from $f^{(1)}_n$,
\beq{}
    \vec{j}^{(3)} =  -e \int_{\kv} \sum_n f^{(1)}_{n} \Bigg( - \frac{e}{\hbar} \Big( \vec{E} \times  [ \nabla_\kv \times (\underline{\underline{\mathcal{G}^n}}(E^2) \vec{E}) ] \Big) \Bigg).
\eeq
On substituting for $f^{(1)}_n$ in terms of $f^{(0)}_n$, and integrating by parts, we have,
\begin{widetext}
\beq{}
    j^{(3)}_\delta = -\frac{e^4 \tau}{3\hbar^3} E_\alpha E_\beta E_\gamma \int_{\kv} \sum_n f^{(0)}_{n} \Bigg(\partial_{k_\delta} \partial_{k_{(\beta}}\mathcal{G}^n_{\gamma \alpha)}(E^2) - 3\Big( \partial_{k_{(\alpha}} \partial_{k_\beta}\mathcal{G}^n_{\gamma)\delta}(E^2) \Big) \Bigg),
\eeq
which decomposes in terms of Ohmic [$\textbf{j}^{(3)} \cdot \vec{E} \neq 0$] and Hall [$\textbf{j}^{(3)} \cdot \vec{E} = 0$] contributions~\cite{Mandal2024},
\beq{}
    (j^{(3)}_\delta)^{\text{H}} = \frac{2e^4 \tau}{3\hbar^3}  E_\alpha E_\beta E_\gamma \int_{\kv} \sum_n f^{(0)}_{n} \Bigg(\partial_{k_\delta} \partial_{k_{(\beta}}\mathcal{G}^n_{\delta \alpha)}(E^2)  -  \partial_{k_{(\alpha}} \partial_{k_\beta}\mathcal{G}^n_{\gamma)\delta}(E^2) \Bigg),
\eeq
\beq{}
    (j^{(3)}_\delta)^\text{O} = -\frac{e^4 \tau}{3\hbar^3}  E_\alpha E_\beta E_\gamma \int_{\kv} \sum_n f^{(0)}_{n} \Bigg( \partial_{k_{(\alpha}} \partial_{k_\beta}\mathcal{G}^n_{\gamma \delta)}(E^2) \Bigg),
\eeq
\end{widetext}
with the corresponding conductivities calculated for the Bloch Euler Hamiltonians in the main text. 

Finally, at third order in electric fields, one could consider the intrinsic Hall conductivity contributions arising from Berry-connection polarizability tensor (BCT) $T^n_{\alpha \beta \gamma}$~\cite{Lai2021, Liu2022},
\beq{}
    (A^{''}_{nn})_\alpha = T^n_{\alpha \beta \gamma} E_\beta E_\gamma,
\eeq
which determines a higher, second-order in electric fields, correction to the Berry curvature,  $\Om^{''}_n = \nabla_\kv \times \vec{A}^{''}_{nn}$. The BCT involves higher-order field-induced perturbative band corrections $\ket{u^{(2)}_{n\kv}}$, and explicitly reads~\cite{Liu2022, Mandal2024}
\beq{}
    T^n_{\alpha \beta \gamma} = \text{Re}~\sum_{m \neq n} \Big(\mathcal{U}_{\alpha \beta \gamma}^{nm} + \mathcal{U}_{\beta \alpha \gamma}^{nm} - \mathcal{U}_{\beta \gamma \alpha}^{nm} - \sum_{p \neq n,m} \mathcal{V}_{\alpha \beta \gamma}^{nmp} \Big),
\eeq
with two-band $\mathcal{U}_{\alpha \beta \gamma}^{nm}$, and three-band $\mathcal{V}_{\alpha \beta \gamma}^{nmp}$ contributions explicitly given by the non-Abelian Berry connection contributions (see also Appendix~\ref{app:A})~\cite{Liu2022, Mandal2024},
\beq{}
    \mathcal{U}_{\alpha \beta \gamma}^{nm} = M^\alpha_{nm} (A^\beta_{nn}-A^\beta_{mm}-i\partial_{k_\beta}) M^\gamma_{mn},
\eeq
\beq{}
    \mathcal{V}_{\alpha \beta \gamma}^{nmp} = (2M^\alpha_{nm} A^\beta_{mp} + M^\beta_{nm} A^\alpha_{mp})M^\gamma_{pn}.
\eeq
The BCT yields intrinsic third-order currents,
\beq{}
    \vec{j}^{(3)} =  -e \int_{\kv} \sum_n f^{(0)}_{n} \Big( - \frac{e}{\hbar} [\vec{E} \times  \vec{\Om}^{''}_n ] \Big),
\eeq
which componentwise can be expressed as~\cite{Xiang2023}
\beq{}
    {j}^{(3)}_\delta = \frac{e^4}{\hbar^3} E_\alpha E_\beta E_\gamma \int_{\kv} \sum_n f^{(0)}_{n} \Big( \partial_{k_\beta} T^n_{\delta \gamma \alpha} - \partial_{k_\delta} T^n_{\beta \gamma \alpha} \Big).
\eeq
However, such intrinsic contributions are vanishing under the considered combination of $\mathcal{C}_2\mathcal{T}$, $\mathcal{C}_2$, and $\mathcal{T}$ symmetries~\cite{Mandal2024}. In the noncentrosymmetric case, we expect these contributions to be dominated by the second-order quantum-metric dipole Hall effect, as long as the electric fields $\vec{E}$ remain moderate, and the scattering time is sufficiently long, i.e., $\frac{|e\vec{E} a_i|}{\hbar} < \frac{1}{\tau}$, where the lattice constants $a_i$ in $i = x,y$ directions physically determine the electric potential differences across the individual unit cells of the system.
\\

\section{Bulk nonlinear currents from continuum models}\label{app:H}

Below, we demonstrate how the bulk nonlinear quantum-metric Hall responses can be directly retrieved from effective continuum models. These findings support the results obtained numerically in the lattice-regularized models hosting patch Euler class topology (Sec.~\ref{sec:III}). As detailed in Appendix~\ref{app:E}, we utilize Christoffel symbols and Riemannian curvature, which we include within the response functions of the corresponding conductivities. To model the inversion-symmetry breaking perturbation (see Sec.~\ref{sec:III}), we modify the original continuum Hamiltonian characterized in Appendix~\ref{app:E} with a perturbation $H_{\Delta} = \Delta k_1 \mathsf{1}$, i.e., the combined Hamiltonian reads,
\beq{}
    H(\kv)=\alpha(k_+^{2|\chi|}\sigma_-+k^{2|\chi|}_-\sigma_+) + \Delta k_1 \mathsf{1},
\eeq
from which we deduce all the nonlinear responses studied in the main text. On computing the integrands of the relevant bulk anomalous Hall conductivities realized in the model with $\chi = 1$, and setting $\alpha=1$, we correspondingly have

\begin{widetext} 
\beq{}
\begin{split}
 \sigma_{y, xx}(\nabla)= \frac{e^2}{\hbar} \intBZ \dd^2 \kv~\theta(3 \Delta k_x + 9 k_x^2 + \sqrt{3} \Delta k_y + 3 k_y^2 - 2 \mu) 
 \theta( 9 k_x^2 + 3 k_y^2 - 3 \Delta k_x - \sqrt{3} \Delta k_y + 2 \mu)\\ \Bigg( \frac{-3 \sqrt{3} k_x^3 + 3 k_x^2 k_y + 
 3 \sqrt{3} k_x k_y^2 + k_y^3)}{3 (3 k_x^2 + k_y^2)^3} \Bigg),
\end{split}
\eeq
\beq{}
\begin{split}
 \sigma_{y, xy}(\nabla)= \frac{e^2}{\hbar} \intBZ \dd^2 \kv~\theta(3 \Delta k_x + 9 k_x^2 + \sqrt{3} \Delta k_y + 3 k_y^2 - 2 \mu) 
 \theta( 9 k_x^2 + 3 k_y^2 - 3 \Delta k_x - \sqrt{3} \Delta k_y + 2 \mu)\\ \Bigg( -\frac{(k_y (-9 k_x^2 + k_y^2)}{3 (3 k_x^2 + k_y^2)^3)} \Bigg),
\end{split}
\eeq
for responses to electric field gradients. For second-order quantum-metric Hall response,
\beq{}
\begin{split}
\sigma_{y, xx}(E^2)= \frac{e^3}{\hbar^2} \intBZ \dd^2 \kv~\theta(3 \Delta k_x + 9 k_x^2 + \sqrt{3} \Delta k_y + 3 k_y^2 - 2 \mu) 
 \theta( 9 k_x^2 + 3 k_y^2 - 3 \Delta k_x - \sqrt{3} \Delta k_y + 2 \mu)\\ \Bigg( -\frac{(-6 (6 + \sqrt{3}) k_x^3 + 12 k_x^2 k_y + 2 (12 + 5 \sqrt{3}) k_x k_y^2 + 
  4 k_y^3}{9 (3 k_x^2 + k_y^2)^4)} \Bigg),
\end{split}
\eeq
\beq{}
\begin{split}
\sigma_{y, xy}(E^2)= \frac{e^3}{\hbar^2} \intBZ \dd^2 \kv~\theta(3 \Delta k_x + 9 k_x^2 + \sqrt{3} \Delta k_y + 3 k_y^2 - 2 \mu) 
 \theta( 9 k_x^2 + 3 k_y^2 - 3 \Delta k_x - \sqrt{3} \Delta k_y + 2 \mu)\\ \Bigg( \frac{18 k_x^3 - 3 (8 + 5 \sqrt{3}) k_x^2 k_y + 
 6 k_x k_y^2 + (4 + \sqrt{3}) k_y^3}{9 (3 k_x^2 + k_y^2)^4} \Bigg).
\end{split}
\eeq
For third-order quantum-metric Hall response,
\beq{}
\begin{split}
\sigma_{y, xxx}(E^3)= \frac{e^4 \tau}{\hbar^3} \intBZ \dd^2 \kv~\theta(3 \Delta k_x + 9 k_x^2 + \sqrt{3} \Delta k_y + 3 k_y^2 - 2 \mu) 
 \theta( 9 k_x^2 + 3 k_y^2 - 3 \Delta k_x - \sqrt{3} \Delta k_y + 2 \mu)\\ \Bigg( \frac{4 (45 (6 + \sqrt{3}) k_x^4 - 108 k_x^3 k_y - 
   6 (51 + 19 \sqrt{3}) k_x^2 k_y^2 - 
   36 k_x k_y^3 + (12 + 5 \sqrt{3}) k_y^4)}{9 (3 k_x^2 + k_y^2)^5}\Bigg),
\end{split}
\eeq
\beq{}
\begin{split}
\sigma_{y, xyx}(E^3)= \frac{e^4 \tau}{\hbar^3} \intBZ \dd^2 \kv~\theta(3 \Delta k_x + 9 k_x^2 + \sqrt{3} \Delta k_y + 3 k_y^2 - 2 \mu) 
 \theta( 9 k_x^2 + 3 k_y^2 - 3 \Delta k_x - \sqrt{3} \Delta k_y + 2 \mu)\\ \Bigg( \frac{32 (36 k_x^4 - 27 \sqrt{3} k_x^3 k_y + 6 k_x^2 k_y^2 + 3 \sqrt{3} k_x k_y^3 - 
   2 k_y^4)}{27 (3 k_x^2 + k_y^2)^5} \Bigg),
\end{split}
\eeq
which functionally capture the characteristic features of the bulk anomalous transport conductivities as a function of chemical potential $\mu$. $\theta$ denotes the Heaviside step function, which imposes the occupations of states over hexagonal BZ in the zero-temperature limit. Correspondingly, we include the continuum model conductivities in Fig.~\ref{Fig:contCond}. Having accounted for the geometric contributions of the third (trivial) band, these results remain in a good qualitative overlap with the anomalous conductance scalings between the Euler band subspace in the lattice-regularized Euler Hamiltonians (Sec.~\ref{sec:III}) which were shown in Fig.~\ref{Fig:Bulk}.  The generalization to an arbitrary Euler class $\chi$ amounts to including $\chi^2$ prefactor contributed by the quantum metric, following the analytical results of Appendix~\ref{app:E}.

\end{widetext}
\bibliography{references}

\end{document}